\documentclass[aps,prd,nofootinbib,preprintnumbers]{revtex4-2}
\usepackage{graphicx}
\usepackage{bbold}
\usepackage{hyperref} 
\usepackage{slashed}
\usepackage{enumerate}
\usepackage{enumitem}
\usepackage{amsmath}
\usepackage{amssymb}
\usepackage{mathrsfs}
\usepackage{MnSymbol}
\usepackage{moresize}
\newcommand{\bL}{\begin{Large}}
\newcommand{\eL}{\end{Large}}

\newcommand{\bea}{\begin{eqnarray}}
\newcommand{\eea}{\end{eqnarray}}
\newcommand{\be}{\begin{equation}}
\newcommand{\ee}{\end{equation}}

\newcommand{\cm}{\color{magenta}}
\newcommand{\cy}{}
\newcommand{\cb}{}
\newcommand{\cbl}{}
\usepackage{tikz-feynman}
\tikzfeynmanset{compat=1.1.0}
\usepackage[margin=1in]{geometry}
\usepackage[colorinlistoftodos]{todonotes}
\usepackage{appendix}

\newcommand{\cP}{\ensuremath{\mathcal{P}}}
\newcommand{\cT}{\ensuremath{\mathcal{T}}}

\newcommand{\cPT}{\ensuremath{\mathcal{PT}}}

\begin{document}

\preprint{\leftline{KCL-PH-TH/2025-{\bf 27}}}

\title{The axion coupling accelerates the Universe through  \cPT-symmetric phases }
\medskip
\author{Leqian Chen$^a$}
\author{Nick E. Mavromatos$^{b,a,c}$}
\author{Sarben Sarkar$^a$}
\medskip 
\affiliation{$^a$Theoretical Particle Physics and Cosmology Group, Department of Physics, King's College London, London, WC2R 2LS, UK}

\affiliation{$^b$ Department of Theoretical Physics and IFIC, University of Valencia and CSIC, E-46100, Valencia, Spain}

\affiliation{$^c$ Currently on leave from: Physics Division, School of Applied Mathematical and Physical Sciences, National Technical University of Athens, Zografou Campus, Athens 157 80, Greece}

\begin{abstract}
 The conjecture  by two of the  authors (N.E.M. and S.S.) that a \cPT-symmetric phase plays  a role in understanding singular renormalisation group (RG) flows for a Chern-Simons (CS) gauge theory of axions, has been reexamined  and significantly improved.  
We have used  the more complete Wetterich equation, which includes gravitational couplings in a systematic way from the start, to understand the emergence of this phase. The singular structure of the RG flows has persisted on including gravitational-couplings, thereby offering further support to the conjecture that \cPT -symmetric phases of (repulsive) gravity characterise string-effective CS gravitational  theories, where the axion is the massless string-model independent axion, which can also play a role of a totally-antisymmetric torsion degree of freedom. This has suggested a novel interpretation of the currently observed acceleration of the expansion of the Universe in terms of such a phase at large (cosmological) scales.  
\end{abstract}

\maketitle

\section{Introduction and Motivation}
 Recently, within the framework of a string-inspired Chern-Simons cosmological model, it was proposed in~\cite{Mavromatos:2024ozk} that the late-time accelerated expansion of the Universe might be understood, not in terms of a small positive cosmological constant, but as a change   from a conventional Hermitian phase of gravity in the ultraviolet to a non-Hermitian \cPT-symmetric phase in the infrared. $\cPT$ symmetry refers to parity (\cP)-~ time-reversal (\cT) symmetry and was first introduced in quantum mechanics and then later in quantum field theory (QFT)~\cite{qft5,Mavromatos:2024ozk,Croney:2023gwy,Ai:2022olh,PTqm1}.\footnote{A brief but comprehensive discussion on the basic properties and features of this symmetry is given in Appendix \ref{sec:PT}, for the benefit of the non-expert readers, so as to place them into context.} 
  The conjecture of \cite{Mavromatos:2024ozk} is based on the presence of singularities in non-perturbative renormalization-group (RG) $\beta$-functions of (massless) axion electrodynamics couplings~\cite{Eichhorn:2012uv} that characterize the low-energy string theory cosmological model under consideration. 
 
 At such singular points of the non-perturbative RG flows, the theory becomes ill-defined in the infrared. The conjecture of \cite{Mavromatos:2024ozk} is that the well-defined theory is a continuation of the Hermitian theory (in the sense of a phase transition)  to a non-Hermitian but well-behaved, \cPT-symmetric theory, which, when coupled to gravity, makes gravity appear repulsive. Upon interpreting the infrared regions, where the singularity in the $\beta$-function occurs, as corresponding to a late cosmological era of the Universe, this theory is conjectured~\cite{Mavromatos:2024ozk} to correspond to the observed late-era  acceleration of an expanding Universe.  In \cite{Mavromatos:2024ozk}, the initial conjecture has been formulated in the context of flat-spacetime axion electrodynamics. The role of gravity, and its RG flow, was considered later in that analysis, and only in the \cPT-symmetric phase, as a 
 supplemental component. It is the purpose of the present work to incorporate gravity right from the start and, thus, verify, in a rigorous way, the conjecture of \cite{Mavromatos:2024ozk}.
 
 In order to understand the details behind
 the conjecture of \cite{Mavromatos:2024ozk}, it is first necessary to review briefly the underlying formalism. In \cite{Mavromatos:2024ozk} we  studied an effective cosmological field theory, termed Stringy Running Vacuum Model (StRVM) ({\it cf.} Appendix \ref{sec:appstRVM}), which, after appropriate compactification to (3+1)-dimensional spacetimes, stems from microscopic string theories at  energies low compared to the string scale; the details of  the compactification  play no essential role in our analysis or conclusions in \cite{Mavromatos:2024ozk} and also here.

The conjecture is based on the presence of singularities in appropriate FRG $\beta$ functions in the Chern-Simons electrodynamics part of \eqref{sea4}. Specifically, the singularities appear~\cite{Eichhorn:2012uv} in the FRG  $\beta$-function of the renormalized coupling $f^{-1}_a$ of the axion (pseudoscalar field) $b(x)$ to the Pontryagin anomaly term~\cite{Eguchi:1980jx} (which we refer to as the {\it axion coupling}):   
\begin{align}\label{pontryagin}
S_{\rm axion-Pontryagin-anomaly}
=
\frac{1}{16\, \pi^2\, f_a}
\int d^4x\;\sqrt{-g}\; b(x)\,
\Bigl(
\;-\;
F_{\mu\nu}\,\widetilde F^{\mu\nu}
\Bigr)\,,
\end{align}
where, for the StRVM, the axion coupling is 
\be\label{faA}
f_a = \frac{1}{16\, \pi^2\,A} = \frac{6}{\pi^2}\,  
\sqrt{\frac{3}{2}}\,\frac{\kappa}{\alpha'} = \frac{6}{\pi^2}\,  
\sqrt{\frac{3}{2}}\,\frac{M_s^2}{M_{\rm Pl}} \,,
\ee
on using the definition of $A$ in \eqref{Adef}. 

The singularities 
in the $\beta$-function of $f_a$ 
appear in the infrared regions of the (renormalised) Hermitian theory, making it ill defined \cite{Eichhorn:2012uv}.  The \cPT~ properties of the set of singularities  and zeros of the beta function \cbl  led to the conjecture that such an ill-defined theory, can be extended (beyond such singularities) to a \cPT- symmetric, non-Hermitian theory. At large cosmological scales (in the far infrared)\cite{Mavromatos:2024ozk}, the $\cPT$ phase  is characterised by regular FRG-dressed axion couplings, but by a {\it repulsive} gravitational interaction, while the short-distance gravity is attractive.\footnote{ In \cite{Ai:2022csx,Chen:2024ynx} potential extensions of ill-defined Hermitian scalar field theories to well-defined \cPT-symmetric non-Hermitian ones
are conjectured,  in a non-gauge setting.  We discuss the extension of this conjecture, in the presence of  gravity, to explain the observed cosmological expansion of the Universe at large scales,  without the need for introducing a dark energy component. We also assume this non-trivial phase change in the infrared region of the curved-space extension of axion electrodynamics \cite{Mavromatos:2024ozk}. 
} There is a parallel  with  \cPT-symmetric  quantum electrodynamics \cite{Milton2011-sd}, where like charges {\it attract} and opposite charges {\it repel}, a feature of \cPT ~symmetry.  

\begin{figure}[ht]
  \centering
\includegraphics[width=0.8\textwidth]{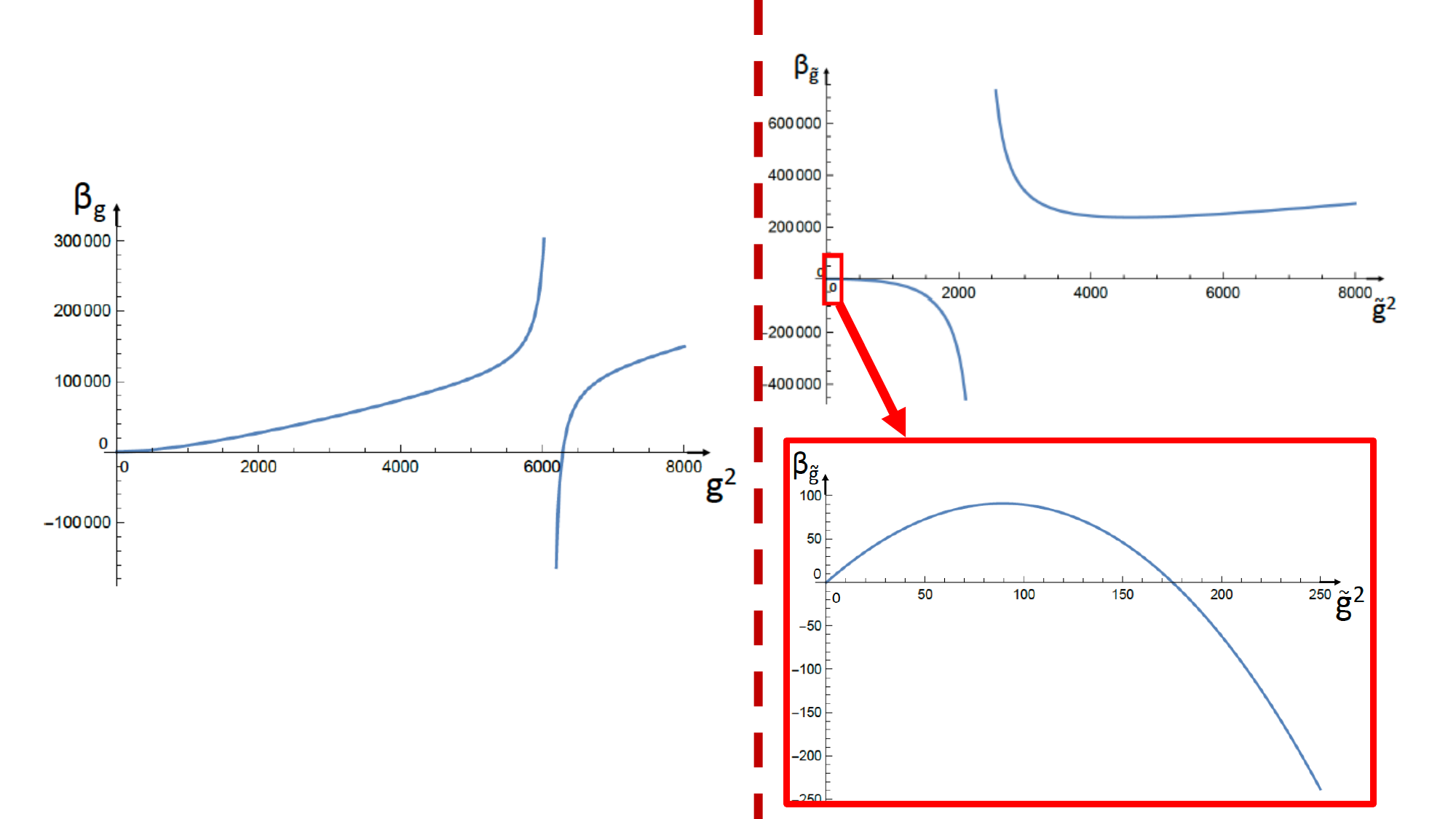}
  \caption{{\it Left panel}: The FRG beta function $\beta_g \equiv \partial_t g^2$ of the Hermitian axion CS electrodynamics theory~\cite{Eichhorn:2012uv}, exhibiting a singular behavior at a finite running coupling $g \equiv (4\pi^2\,f_a)^{-1}$, as well as a fixed point behavior, characterised by the existence of a trivial fixed point at zero coupling, and a non-trivial one at strong coupling, $g_\star^2 \gg 1$. {\it Right Panels}: Nonperturbative RG beta function for non-Hermitian case as a function of $\tilde g^2$, where $\tilde g = i g$.
  The behaviour is qualitatively different from the Hermitian beta function~\cite{Eichhorn:2012uv}. The area included in the red rectangle includes a nontrivial fixed point (apart from the trivial one at zero coupling), which is indicated explicitly in the lower  left panel.
  Figure synthesized from relevant graphs in \cite{Eichhorn:2012uv} and \cite{Mavromatos:2024ozk}.}
  \label{fig:conjecture}
\end{figure}

 As already mentioned, in \cite{Mavromatos:2024ozk} the emergence of the $\cPT$ phase was first investigated in a flat background. The situation is depicted concisely in figure~\ref{fig:conjecture}. There is a singular behaviour of the FRG  $\beta$-function of the axion coupling $f_a$ in the infrared, making the theory ill-defined at large distances (left panel). Thus, it is not possible, by staying within the Hermitian theory, to connect the interacting theory with the free theory at the trivial fixed point $g=0$ smoothly, via an appropriate RG flow from the ultraviolet (UV) to the infrared fixed points.

 In  \cite{Mavromatos:2024ozk} this is remedied by conjecturing that the theory 
 transitions (at the running RG energy scale $k=k_{\rm sing}$) to its non-Hermitian counterpart, obtained by analytic continuation of the coupling 
\be\label{analcont}
f_a \to i f_a\,.
\ee
which exhibits a well-defined regular behaviour in the infrared~\cite{Eichhorn:2012uv}.
The non-Hermitian theory is characterised by a singularity in the (trans)UV regime, a non-trivial UV fixed point, and a smooth running from the UV to the trivial infrared (attractive) fixed point at $\widetilde g=0$ ({\it cf.} right panels of figure \ref{fig:conjecture}). 
Here we followed the logic in \cite{Romatschke:2022llf} for ordinary Landau poles,
inspired by the conjecture of \cite{Ai:2022csx},
whereby the singularity is bypassed by an analytic continuation of the theory. The analytic continuation of the axion coupling $f_a$ is a necessary condition for \cPT ~symmetry, as follows from the discussion in \cite{Ai:2022olh}, for scalar field theories.   

In \cite{Mavromatos:2024ozk},
it has been stressed that Eq.~\ref{faA}, which is valid {\it only} for string-model-independent (or KR) axions ({\it cf.} Appendix \ref{sec:appstRVM}), 
is {\it crucial} for the validity of the conjecture of \cite{Mavromatos:2024ozk}.  Indeed, the latter relation implies that, upon the analytic continuation 
\eqref{analcont}, which allows for the transition to the non-Hernitian theory, the gravitational coupling 
\be\label{analcontgrav}
\kappa \, \to \, i\, \tilde \kappa 
\ee
while the square of the string scale remains real.\footnote{Even if one analytically continued $M_s$, $M_s^2$ remains real ($M_s^2 \in \mathbb R$). However, the physically correct approach is the one in which one does not analytically continue $M_s$, due to  
the requirement that the compactification radii in string models, which are proportional to $M_s$, must remain real~\cite{Mavromatos:2024ozk}.}
In view of \eqref{analcontgrav} one, then, obtains, as explained in \cite{Mavromatos:2024ozk}, a repulsive phase of gravity for the analytically continued phase of the axion electrodynamics model, when placed in a gravitational background:
\begin{align}\label{gtilde}
 G = \frac{\kappa^2}{8\pi} \quad \rightarrow \quad  \widetilde G = - \frac{{\widetilde \kappa}^2}{8\pi}\, < \, 0 \,,
 \end{align}
where $G$, $\tilde G$ denote the respective Newton constants in the Hermitian and $\cPT$~symmetric phases, respectively.

However, the alert reader might raise objections to the above treatment due to the fact that 
in the above analysis, we started from the flat-spacetime singularities of axion electrodynamics, and 
only later, within a $\cPT$- symmetric phase, is the renormalisation flow of gravity considered  as a supplemental component.
The RG analysis of \cite{Mavromatos:2024ozk} did not incorporate quantum gravitational flows from the start and so  did not take into account the back reaction 
 of gravitational fluctuations.  
 That analysis was based on a Wetterich equation without gravity~\cite{Wetterich:1992yh,Berges:2000ew}. Hence the next step is the incorporation of gravity  in a Wetterich formulation.
This is the basic purpose of the current work, namely to incorporate the role of gravity from the start, and examine its implications for the Landau-pole behaviour of the dynamical Abelian gauge CS coupling and $\cPT $ symmetry, crucial for our earlier analysis. We shall demonstrate that the singularities in the FRG flow of the axion coupling, that led to the conjecture of \cite{Mavromatos:2024ozk}, remain intact after the inclusion of gravity.

 At this stage we would like to mention an important point. As discussed above, in the context of the StRVM, which motivates the work, there are no uncancelled gCS anomaly terms of the form \eqref{e1} at post-RVM-inflationary eras at large scales, in particular the late cosmic eras characterised by an accelerated expansion. 
Thus they will be ignored in this work.\footnote{ Their presence would complicate the FRG equations in a way explained briefly in subsection \ref{sec:gravanomint}.} 
From the perspective of perturbative quantum fields in four dimensions, the theory \eqref{sea1} is non-renormalisable, Thus, in \cite{Mavromatos:2024ozk} and here, we are forced to use a nonperturbative formulation, such as the one proposed by Wetterich~\cite{Wetterich:1992yh,Berges:2000ew}, for the functional renormalization group (FRG) framework~\cite{Reuter:2019byg,Dupuis:2020fhh},
and revisit crucial non-renormalisation arguments on topological grounds.

The structure of the article is the following:  Section \ref{sec:FRG} sets up the non-perturbative flow equations for Euclidean effective actions used in our FRG analysis of axion electrodynamics with gravity, generalising \cite{Mavromatos:2024ozk}. As mentioned in the introduction, this model is an essential part of the StRVM at late epochs~\cite{bms,ms1}, which are of interest to us here. Section \ref{sec:gravaxion} introduces dynamical gravity into axion electrodynamics; within the StRVM framework we omit gCS anomaly terms. Section \ref{sec:gravbetaanomdim} computes the gravitational contributions to the FRG $\beta$ functions and anomalous dimensions of axion electrodynamics and  presents our main results: under reasonable, physically motivated approximations ({\it cf.} \ref{sec:results}), the Wetterich equation leaves the $\beta$-function singularities of the axion coupling found in \cite{Eichhorn:2012uv} intact even in the presence of gravity ({\it cf.} \ref{sec:LPresult}), thereby supporting the conjecture of \cite{Mavromatos:2024ozk} that gravity may exhibit a repulsive phase at large scales. In subsection \ref{sec:stringpheno}, we give, for completion, some elementary estimates on the magnitudes of couplings, within the context of our string-inspired model.
Finally, Section \ref{sec:concl} presents our conclusions and outlook. 
Appendix \ref{sec:appstRVM}
reviews the essential elements of StRVM, which the conjectured of \cite{Mavromatos:2024ozk} is based upon.  Appendix \ref{sec:PT} reviews briefly the basic features of 
\cPT~symmetry, mainly for introducing the non expert reader to the subject. Appendix \ref{sec:PTCS} reviews the \cPT-symmetric aspects of non-Hermitian Chern–Simons (CS) theories that underpin our approach.
Appendix  \ref{sec:PTanomcanc}
derives the emergence of CS terms from anomaly-cancellation arguments and clarifies that the string-inspired reasoning of \cite{Mavromatos:2024ozk} extends beyond string theory. Appendix \ref{sec:analcontstring} further motivates the \cPT-symmetric framework~\cite{mavromicro} by addressing a known ambiguity in the appearance of CS gravity in (bosonic) string theory upon analytic continuation from Minkowski to Euclidean signature; this step is required to define a well-posed path integral for the low-energy gravitational effective action.  Appendix \ref{sec:singflows}
reviews simple examples of singular perturbative renormalization-group  \cite{, Bender:2021fxa} in field theory, where situations like the one conjectured in \cite{Mavromatos:2024ozk} might occur, but much  simplified and in the absence of gravity. Its purpose is to demystify, as much as possible, the situation described in \cite{Mavromatos:2024ozk}.
 Appendices \ref{EffAction} and\ref{sec:FRGscalar}  introduce and illustrate key features of the FRG in a flat-space scalar toy model, introducing elements of the Wetterich formalism and the role of topology in constraining exact RG flows.
Appendices \ref{DiffForms}-\ref{LPAModified} summarise terminology and technical aspects of the non-perturbative FRG framework, which we make use of in the main text.

\section{The functional renormalisation group (FRG)}\label{sec:FRG}

The  Wilsonian renormalisation group (RG)~\cite{Wilson:1971bg,Wilson:1973jj} provides a framework for understanding the scale dependence of quantum field theories. In RG, one integrates out quantum fluctuations in momentum shells from a UV scale $\Lambda$ down to a lower scale $k$, thereby constructing a sequence of scale-dependent effective actions $S_k(\Phi)$. Couplings evolve as high momentum modes are eliminated. This procedure usualy requires a sharp momentum cut-off, which can obscure gauge and diffeomorphism invariance. It is also cumbersome since it requires frequent reparametrisations and is not framed in terms of the one-particle irreducible (1PI) effective action. By contrast the Wetterich equation \cite{Wetterich:1992yh} evolves the effective \emph{average} action $\Gamma_k[\Phi]$ - a scale-dependent version of the 1PI effective action- which includes the effects of quantum fluctuations with momenta $p^{2} \gtrsim k^{2}$ while suppressing those below $k^2$ via a smooth regulator function $R_{k}\left(p^{2}\right)$ (see Appendix \ref{EffAction}).  This formulation (an example of the functional renormalisation group (FRG))  remains exact even in strongly coupled regimes; it allows systematic truncations, which are approximations that go beyond perturbation theory (e.g. derivative expansions) that retain nontrivial quantum information, while being computationally tractable. 

The effective average action satisfies the  Weterrich equation~\cite{Reichert:2020mja,Wetterich:1992yh,Eichhorn:2012uv,Eichhorn:2018yfc,deBrito:2022vbr}
\be
\label{Wetterich_1}
\partial_{k} \Gamma_{k}[\Phi]=\frac{1}{2}  ST{r}\left[\left(\Gamma_{k}^{(2)}[\Phi]+R_{k}\right)^{-1} \partial_{k} R_{k}\right]
\ee
where $\Phi$ denotes the fields of the theory (e.g. axions, gauge fields, metric); $\Gamma_{k}^{(2)}$ is the second functional derivative (Hessian) of $\Gamma_k$ with respect to $\Phi$; $R_{k}$ is an infra-red regulator function that introduces a momentum dependent mass to suppress fluctuations with $p^{2} \lesssim k^{2}$. STr is a trace that includes both a functional trace over momenta and an internal trace over field indices. All our analysis will be based on this  framework (see Appendix \ref{EffAction}).

Although \eqref{Wetterich_1} resembles a one-loop formula \cite{Reichert:2020mja}, all  vertices and propagators are fully dressed (coming from $\Gamma_{k}^{\left( 2 \right)}$) so no approximation is made. Boundary conditions are implemented via the limiting behaviours of $R_k$. In order to make the approach tractable approximations need to be made. General relativity and its modifications are conventionally non-renormalisable theories but still predictive for  $E\ll M_{Pl}$. Because of non-renormalisability quantum gravity generates an infinte tower of higher-dimensional operators (as noted in the effective field theory (EFT) point of view). If one tries to absorb loop corrections into a single running of the Newtonian coupling constant, after making truncations of the operators mixing under renormalisation, there is typically no unique, universal $\beta$ function \cite{Anber:2010uj}. We will not deal with the difficult issue of truncation dependence of beta function fixed points. Such issues are related to a  full theory of quantum gravity; our considerations do not claim an understanding of this issue.

The universal content of FRG lies in the fixed point values and critical exponents of dimensionless ratios such as $G k^2$ \cite{Eichhorn:2018yfc}. In asymptotic safety it is fixed points controlling the $k \to \infty$ behaviour that is studied \cite{Eichhorn:2018yfc}. Another approach \cite{Anber:2010uj,Ellis:2010rw,Ellis2011-td} (in terms of on-shell couplings) is more suited for $k \to 0$. The two approaches complement each other \cite{Eichhorn:2018yfc}.
Our purpose is to consider topological gravitational theories, which are not usually considered within the FRG framework; there are symmetry considerations which lead to non-renormalisation of couplings. This non-renormalisation is  important in extending the analysis in \cite{Mavromatos:2024ozk}. Our calculations are based on those in \cite{Eichhorn:2012uv,PhysRevD.86.105021,deBrito:2022vbr,Dona:2015tnf,Dona:2015tnf}; however the effect of the gravitational CS term on the axion is not required explicitly following arguments in \cite{bms,ms1,ms2} and discussed in  Appendix \ref{sec:appstRVM}. In order to familiarise the general reader with the machinery of the functional renormalisation group we will first examine a toy example for a single component scalar field theory in Appendix \ref{sec:FRGscalar}; we  indicate how the ideas extend to more general theories of \eqref{Wetterich_1}.The Appendix  introduces the modified local potential approximation (LPA$^\prime$)  \cite{Hellwig:2015woa}, which we  use in our Chern-Simons theory. LPA$^\prime$ is the simplest framework incorporating wave-function renormalisation, essential for discussing anomalous dimensions. It illustrates the calculation  of the running of the anomalous dimensions and couplings  of the renormalisation in terms of threshold functions, a feature  common to the more complex theories that we consider.

\subsection{Renormalisation and  topology}
Topology and symmetry constrain the solutions of the Wetterich equation in gauge theories. The language of differential forms (briefly summarised in Appendix \ref{DiffForms}) is useful in topological settings \cite{Darling1994-lw} and allows a more compact exposition without the need for  explicit factors of the Levi-Civita tensor and (torsionless) metrics. A part of our action coincides with \emph{axion electrodynamics} (written in terms of a $U(1)$ gauge field with a field strength 2-form $F$ and a 0-form axion field $b$); the associated average 1PI action is
\be
\label{Avgaction_1}
\Gamma_{k}=\int_{M}\left\{\frac{1}{2} Z_{A, k} F_{\wedge} * F+\frac{1}{2} Z_{b, k} d b \wedge * d b+\frac{\cy g_{\text{mix},k}\cbl}  {4} b F\wedge F-V(b) *1\right\},
\ee
(and equivalently in index notation
\be
\Gamma_k
=
\int d^4x\,\sqrt{g}\,
\left[
\frac{Z_{A,k}}{4}\,F_{\mu\nu}F^{\mu\nu}
+
\frac{Z_{b,k}}{2}\,\partial_\mu b\,\partial^\mu b
+
\frac{g_{\text{mix},k}}{4}\,
b\,F_{\mu\nu}\tilde F^{\mu\nu}
-
V(b)
\right]),
\label{eq:axionED_index}
\ee
where, for generality, we have included a potential $V(b)$ (which will be taken to be zero on requiring shift-symmetry,implied by our string-theory based framework). For our full system, there are additional contributions $\Delta \Gamma_{k}$: the gravitational CS term \footnote{This term, because of StRVM, does not contribute to our model \cite{bms,bms2,ms1,ms2}, but is included in this discussion in case one wishes to consider more general CS gravity models.}, the dynamical CS term with the axion coupling and the Einstein-Hilbert action. Explicitly 
\be
\label{Avgaction_2}
\Delta \Gamma_{k}=\frac{i \theta_{k}}{384 \pi^{2}} \int_{M} R^{a c} \wedge R_{a c}+\frac{i \cy {g}_{N, k} \cbl }{192 \pi^{2}} \int_{M} b R^{d c} \wedge R_{dc}-\frac{1}{16 \pi G_{k}} \int_{M} R *1
\ee
(and equivalently in index notation
\begin{equation}
\Delta\Gamma_k
=
\int d^4x\,\sqrt{g}\,
\left[
\frac{i\theta_k}{384\pi^2}
R_{\mu\nu\rho\sigma}\tilde R^{\mu\nu\rho\sigma}
+
\frac{i g_{N,k}}{192\pi^2}
b\,R_{\mu\nu\rho\sigma}\tilde R^{\mu\nu\rho\sigma}
-
\frac{1}{16\pi G_k}R
\right]).
\label{eq:gravCS_index}
\end{equation}

Here the curvature 2-form in the orthonormal frame is $R_{b}^{a}=d\omega _{b}^{a}+\omega^{a}_{c}\wedge {\omega} ^{c}_{b}$ (where the one form $\omega^{a}_{c}$ is the spin connection) or equivalently $R_{b}^{a}=\dfrac{1}{2}R_{b\mu \nu }^{a}dx^{\mu }\wedge dx^{\nu}$ in terms of the Riemann tensor; $R$ is the Ricci scalar. 

For all gauge fields (including gravity) we also need to add \emph{gauge fixing} terms $\Delta \Gamma_{G F} $  and \emph{Fadeev-Popov ghost} terms in order to eliminate nonphysical degrees of freedom but they are not relevant for our arguments based on topology. Furthermore, to avoid overburdening our exposition,   we do not explicitly consider non-Abelian gauge fields since our analysis extends to this case in a straightforward way.

 In the absence of $V(b)$ the effective action is invariant under the constant shift $b(x) \to b(x)+\zeta$ since $\int bR\wedge R\rightarrow \int \left( b+\zeta \right) R \wedge R$ and $\int bF\wedge F\rightarrow \int \left( b+\zeta \right) F \wedge F$; the resulting change is $\zeta \int (F\wedge F+R \wedge R)$. Owing to the topological nature of these terms
 \be
 \zeta \int F\wedge F=\zeta \int d\left( A\wedge F\right) =0
 \ee
 and 
 \be
  \zeta \int R \wedge R= \zeta \int d\left( \omega \wedge R-\dfrac{1}{2}\omega \wedge \omega \wedge \omega \right) =0.
  \ee
We use here the general result that for an oriented $d$-dimensional manifold $M$ (with boundary $\partial M$) and $u \in \Omega^{d-1}(M)$, we have $\int _{M}du=\int _{\partial M}u$. If $u$ vanishes on ${\partial M}$ then $\int _{M}du=0$. In our case the boundaries of our spacetime manifold is at  infinity and our fields are assumed to fall off sufficiently fast for boundary contributions to vanish.

This shift symmetry has implications  for renormalisation of the relevant coupling constant $\tilde g$ (a generic notation, not to be confused with gravitational metric or coupling). If $g^{CS} \to g^{CS}+\delta \tilde g$ under renormalisation we can introduce a multiplicative field renormalisation:
\be
b(x)= \zeta b'(x)
\ee
where $\zeta=g^{CS}/(g^{CS}+\delta \tilde g)$ so that
\[
\left( \tilde g+\delta \tilde g\right) \int bR\wedge R=\tilde g\int b'R\wedge R.
\]
Hence
\be
Z\left[\tilde g+\delta \tilde g \right]=\int Db e^{-S\left[ b;\tilde g+\delta \tilde g\right] }=\int Db'e^{-S\left[ b';\tilde g\right] }=Z\left[\tilde g \right],
\ee
which implies that $\delta \tilde g=0$, i.e. $\tilde g$ is not renormalised. As noted in \cite{Eichhorn:2012uv} the physical observables are in terms of $ g_{\text{mix},R}^{2}=\dfrac{ g_{\text{mix},k}^{2}}{Z_{A}^{2}Z_{b}}$ and the running of $\cy g_{\text{mix},R}\cbl$ is due to  running of $Z_{A}$ and $Z_b$, but \emph{not} $\cy g_{\text{mix},k}\cbl$. A similar analysis implies the nonrenormalisation of the gravitational (and any nonAbelian)  CS coupling. 

 The above argument relies on shift symmetry and is appropriate for our case since the axion does not have a potential term (including a mass term). It is possible to relax this requirement \cite{Eichhorn:2012uv} by using $U(1)$ gauge invariance and Bose symmetry, which allows us to write the full $3$-point function as
\be
\Gamma_{k,\  bAA}^{\mu \nu} \left( p_{b},p_{1},p_{2} \right) =i g_{\text{mix},k} \,\mathcal F_{k}\left( s,t,u \right) \epsilon^{\mu \nu \rho \sigma} p_{1\rho}p_{2\sigma}
\ee
where, in the from factor $\mathcal F_{k}$, $s,t$ and $u$ are the Mandelstam variables defined by
\[ s=\left( p_{1}+p_{2} \right)^{2} ,\  \  t=(p_{2}+p_{b})^{2},\  \  u=\left( p_{b}+p_{1} \right)^{2}.\]
In a derivative expansion (at the basis of our effective action) any $\mathcal F_{k}$ is analytic at small external momenta, i.e. $F\left( s,t,u \right) =F_{0}+O\left( p^{2}/k^{2} \right)$ for generic external momentum $p$. From our previous topological arguments any $F_0$ generated through higher order loops gives a vanishing contribution to the running of $ g_{\text{mix},k}$.

\subsection{Comments on the incorporation of the axion-gravitational-anomaly interactions}\label{sec:gravanomint}
The StRVM model of cosmology  motivates the approach in \cite{Mavromatos:2024ozk}. In view of the cancellation of the gCS anomalies in post RVM-inflationary eras of the Universe, 
such terms have not been incorporated in our FRG analysis. 
In this subsection we briefly discuss the potential role of gCS anomalous interactions with the KR axion, as far as the aforementioned FRG flows are concerned. In view of the increase complecity, and the aforementioned motivation within the StRVM framework, we shall not present any analysis of such terms here, but we shall outline potentiual ways of dealing with them, in more general situations in which they are present. 

 First of all we remark that, in general, the inclusion of non-perturbative FRG methods to treat the renormalization of the coefficients of the gravitational and gauge CS terms in \eqref{e1} is expected to lead to different FRG flows for these two couplings. This is a first issue of concern, given that the anomalous terms \eqref{e1}
 are already quantum effects arising from the addition of appropriate RG counterterms in string theory, to ensure cancellation of gauge and gravitational anomalies. 
 As such, in a full string theory should be exact. 
 This is the meaning of the introduction of the Lagrange multiplier KR field $b$ in the effective string-inspired  path integral, to implement the Bianchi constraint \eqref{modbianchi2}, following the exact redefinition \eqref{csterms}
 of the antisymmetric-tensor field strength, $H_{\mu\nu\rho}$, as discussed in the introduction. Truncating the full string theory to a gravitational field theory, and using a QFT, rather than string theory, non-perturbative FRG, based on the Wetterich equation \eqref{Wetterich_1}, is therefore not fully appropriate for the string-inspired case at hand. 
 
 If the gCS interaction in the Wetterich truncated formalism is characterised by finite FRG behaviour of the pertinent coupling in the analytically-continued \cPT-symmetric theory, in a similar manner to the axion coupling in the flat-spacetime axion electrodynamics~\cite{Eichhorn:2012uv}, then the conclusions of \cite{Mavromatos:2024ozk} remain intact, without the need to pereform an explicit string theory extension of the Wetterich formalism, which is currently not available. If, on the other hand, the contribution of the gravitational CS terms is such that the \cPT symmetric theory is characterised by a singular FRG flow of  the gCS coupling in the infrared regime, then the work of \cite{Mavromatos:2024ozk} will be invalidated, since, in contrast to the flat case, the gravitationally anomalous theory would still be characterised by singular flows, and thus the continuation of \cite{Mavromatos:2024ozk} would be meaningless within this truncated field-theoretic FRG approach.

 To ensure the string-theory considerations of Green-Schwarz mechanism~\cite{Green:1984sg} are respected by our truncated FRG analysis, we propose below to impose the constraint that the CS terms \eqref{e1} remain intact under FRG, that is,  the couplings of the gCS and gauge CS anomaly terms are renormalized in the same way, so their FRG flows remain the same, thus reducing the dimensionality of the coupling space. In other words, one is compelled to use an appropriatelly {\it constrained Wetterich equation} to tailor the FRG to our string-inspired system \eqref{e1}.

  For completeness we note that a similar argument~\cite{DeWitt:1967ub} can be made for the nonrenormalisation of the coupling for $bR\tilde R$ in the presence of a potential for $b$. We shall discuss gravity and its quantisation later. The 1PI vertex for one axion and two gravitons (after gauge fixing and TT projection) is 
\be
\Gamma_{bhh}^{\mu \nu ,\rho \sigma} \left( p_{b},p_{1},p_{2} \right) =i \cy {g}_{N, k} \, \epsilon^{\alpha \beta \gamma \delta} p_{1\alpha}p_{1\beta}p_{2\gamma}p_{2\delta}T^{\mu \nu ,\rho \sigma}\left( p_{1},p_{2} \right) F\left( s,t,u,... \right)
\ee
where $T^{\mu \nu ,\rho \sigma}$ is rank-2 transverse projector such that 
$p_{1\mu}T^{\mu \nu ,\rho \sigma}=0$ or $p_{2\rho}T^{\mu \nu ,\rho \sigma}=0$ (cf. \eqref{eq:gravCS_index}). The diffeomorphism and local Lorentz Ward identity \cite{DeWitt:1967ub} requires the vertex to be transverse in each graviton leg. In a derivative expansion $F=F_{0}+O\left( p^{2}/k^{2} \right)$ which is related to a constant $b$ and by our previous arguments there is no running of the coupling constant. For phenomenological reasons (based on condensates in the  running vacuum model \cite{rvm1}) this term involving the Pontryagin density will not be part of our minimal model for including the effect of gravity.

\subsection{Dependence of the Axion Anomalous Dimension \texorpdfstring{\(\eta_ b \)}{}  on the Photon Anomalous Dimension \texorpdfstring{\(\eta_A\)}{} in Axion Electrodynamics}\label{sec:axanphotan}
The classical action for axion electrodynamics is:
\begin{equation}
S = \int d^4x \left[
  \frac{1}{4} Z_{A} F_{\mu\nu} F^{\mu\nu}
  + \frac{1}{2} Z_b \partial_\mu b\, \partial^\mu b
  + \frac{\cy g_{\text{mix},k}\cm}{4} \, b  F_{\mu\nu} \tilde{F}^{\mu\nu}
\right];
\end{equation}
we functionally differentiate the Wetterich equation
twice with respect to the  axion fields $b(p)$, $b(-p)$ (in Fourier space), and project onto the kinetic term (i.e., isolate the coefficient of \(p^2\)).
The term $b F \tilde{F}$ generates a vertex between one axion and two photons. Therefore, one-loop diagrams contributing to $\eta_b$ involve internal photon lines,
where 
\be\label{anomdimAb}
\eta_{b (A)} = - \frac{\partial}{\partial t}Z_{b(A)}\,, \qquad t \equiv \ln k\,,
\ee
denote the corresponding quantum field $b$ or 
Even in absence of a direct axion loop, photon dynamics can induce non-trivial flow for $\eta_b$. The structure is universal for dimension-5 operators of the type $b F \tilde{F}$, and can be seen in general settings in axion electrodynamics: $\eta_b$ sees $\eta_{A}$ via the loop diagrams generated by the $b F \tilde{F}$ interaction. This coupling manifests in FRG through $\partial_t R_k^A$, which carries $\eta_{A}$.The result is non-perturbative, and essential for accurately describing the renormalization of axion–photon systems~\cite{Eichhorn:2012uv}:

\begin{align}
 \eta_{b}&=\dfrac{ g_{\text{mix}}^{2}}{6\left( 4\pi \right) ^{2}}\left( 2-\dfrac{\eta _{A}}{4}\right) \label{nonPert1}\,,
 \\
\eta_{A}&=\dfrac{ g_{\text{mix}}^{2}}{6\left( 4\pi \right) ^{2}}\left( 4-\dfrac{1}{4}\left( \eta _{b}+\eta _{A}\right) \right) \,.\label{nonPert2}  
\end{align}
In recent work \cite{Mavromatos:2024ozk, Eichhorn:2012uv}  the renormalisation flow of the coupling \footnote{Note that as we shown above and stated in \cite{Mavromatos:2024ozk,Eichhorn:2012uv}, $g_{\text{mix},k}$ is non-renormalisable, and hence not dependent on the energy scale $k$, while $g_{\text{mix}}$ gain dependence on the  energy scale $k$ by the flow of the fields.}
\be\label{gmixdef}
\cy g_{\text{mix}}^{2}=\frac{\cy g_{\text{mix},k}^{2}k^{2}}{Z_{A}^{2}Z_{b}}
\ee
has been investigate. In the massless $b$ case the flow of $g_{\text{mix}}$ is given by 
\begin{equation}
\label{RG_flow}
\partial_t  g_{\text{mix}}^2
= 2 g_{\text{mix}}^2\,
\frac{
  13 g_{\text{mix}}^4
  - 8064 \pi^2 g_{\text{mix}}^2 
  - 147456 \pi^4 
}{g_{\text{mix}}^4
  - 384 \pi^2 g_{\text{mix}}^2 
  - 147456 \pi^4 
}.
\end{equation}
The Landau pole structure of this flow led to speculation connecting Hermitian theory to \cPT-symmetric theory, which led to the possibility of a repulsive gravity phase in the \cPT-symmetric theory \cite{Mavromatos:2024ozk}. Since gravity was not included in the theory which showed the the interesting features at the Landau pole, our purpose it to investigate whether the \emph{Landau pole }features remain \emph{when gravity is included}.

\section{Introduction of gravity into the axion model}\label{sec:gravaxion}
 In a four dimensional spacetime the quantum theory of gravity is highly nontrivial; as a conventional field theory it is non-renormalisable and  needs to respect the exact Ward identities of diffeomorphism invariance. This sets it out from the scalar theory discussed in \ref{sec:FRGscalar}. In the FRG approach we integrate out field fluctuations by wavelength. To discuss wavelengths we need a background metric denoted by $\overline g$ (which can be flat). The FRG cuts off modes with momentum below a scale $k$. For a tensor field, like the spacetime metric, momentum is defined using a Laplacian built using the background. Gravity has gauge redundancy (diffeomorphisms). To quantise the theory, as is common to gauge theories, it is necessary to gauge fix  and add ghosts (using a background connection). We have two options, one being to use a (slightly) curved background (and the heat kernel method \cite{Vassilevich:2003xt}) to project to terms (in the effective action) or use a flat background but extract couplings from fluctuation verices (e.g. $2-$ and $3-$ point functions at some external momentum). Implementing gauge invariance consistently in approximate formulations is one of the difficulties of the method. The fluctuation $h_{\mu\nu}$ around the background decomposes into spin-$2$, spin-$1$ and spin-$0$ parts (irredicble decompositions with respect to $SO(3)$ \cite{Maggiore:2007ulw}).If we perform the York decomposition of the symmetric tensor fluctuation $h_{\mu\nu}$ around a fixed background metric $\bar{g}_{\mu\nu}$ and decompose the full metric $g_{\mu\nu}$, any symmetric tensor fluctuation $h_{\mu\nu}(x)$ can be uniquely decomposed as ({\it cf.} Appendix \ref{FlatYork}):
\begin{equation}\label{eq:York}
  h_{\mu\nu}
  \;=\;
  h_{\mu\nu}^{\mathrm{TT}}
  \;+\;
  \bar{\nabla}_{\mu}\,\xi_{\nu}
  \;+\;
  \bar{\nabla}_{\nu}\,\xi_{\mu}
  \;+\;
\Bigl(\bar{\nabla}_{\mu}\bar{\nabla}_{\nu} - \tfrac{1}{4}\,\bar{g}_{\mu\nu}\,\bar{\nabla}^{2}\Bigr)\,\sigma
  \;+\;
  \tfrac{1}{4}\,\bar{g}_{\mu\nu}\,h,
\end{equation}
where: $h_{\mu\nu}^{\mathrm{TT}}$ is the \emph{transverse‐traceless (TT)} tensor component, $h$ and $\sigma$ are scalars and  $\xi_{\mu}$ is a vector.  In a rigorous treatment the renormalisation of these different pieces of the metric are connected owing to diffeomorphism invariance.\footnote{Another use of such a decomposition is to monitor the efficiency of an approximate calculation in maintaing diffeomorphism invariance \cite{Dona:2015tnf}.}.   Using a background (including a flat one ) is not about claiming that spacetime has really that background. The background is a useful scaffold \cite{Reichert:2020mja} which organises gauge symmetry and the background-covariant regulator gives a soft, gauge-compatible infrared mass to all spin sectors (and so avoids infrared divergences).

The metric is treated in the background-field formalism with a linear split:
\be\label{backexp}
g_{\mu\nu} = \overline g_{\mu\nu} + \kappa h_{\mu\nu},\qquad \kappa = \sqrt{32\pi G_N}\,.
\ee
Ignoring Chern-Simons terms in \eqref{Avgaction_2} for the moment, the gravity ansatz below is represented by an Einstein–Hilbert action, background-covariant gauge fixing term and Fadeev-Popov ghosts\cite{Reuter:2019byg,Codello:2008vh}
\be
\label{Avgaction_2a}
\Gamma_k^{\text{grav}} = -\frac{1}{16\pi G_N} \int\! d^4x \,\sqrt{g}\,R + \frac{1}{2\alpha_{\rm gf}} \int\! d^4x \,\sqrt{\bar g}\,\bar g^{\mu\nu}\, F_\mu[h;\bar g]\,F_\nu[h;\bar g] + \Gamma_k^{\text{ghost}}\ee
where the gauge-fixing functional is chosen as
\begin{equation}
  F_\mu[h;\bar g]
  =
  \Big(
    \delta^\alpha_\mu \,\bar g^{\nu\beta}
    - \frac{1+\beta_{\mathrm{gf}}}{4}\,\delta^\nu_\mu \,\bar g^{\alpha\beta}
  \Big)\,
  \bar\nabla_\nu h_{\alpha\beta}.
\end{equation}
Here $\bar\nabla$ is the covariant derivative with respect to the background
metric $\bar g_{\mu\nu}$, and $\alpha_{\mathrm{gf}}$, $\beta_{\mathrm{gf}}$
are gauge parameters. In the ``standard gravity'' analysis we consider the
Landau (de Donder) gauge,
\begin{equation}
  \alpha_{\mathrm{gf}} \to 0,
  \qquad
  \beta_{\mathrm{gf}} = 0.
\end{equation}
In this gauge, the TT and a scalar  term in the decomposition remain. $\Gamma_k^{\mathrm{ghost}}$ is the Faddeev--Popov ghost action
obtained from the gauge-fixing condition $F_\mu$.

In this truncation we do \emph{not} include an explicit cosmological constant
term, in order to keep the setting close to perturbative gravity about flat space. Gauge fixing $F_\mu[h;\bar g]$ and the ghost action $\Gamma_k^{\mathrm{ghost}}$
        are constructed to be \emph{covariant under background diffeomorphisms},
        i.e.\ transformations generated by a vector field $\xi^\mu$ acting as
        $\delta \bar g_{\mu\nu} = \mathcal{L}_\xi \bar g_{\mu\nu}$, with corresponding
        transformations on $h_{\mu\nu}$ and the matter fields. In this way, the effective average action $\Gamma_k$ is invariant under
\emph{background} diffeomorphisms, while the full diffeomorphism (``split'')
symmetry between $g$ and $\bar g$ is broken by the gauge fixing and by the IR regulator $R_k$. The contributions of the various bits of the decomposition can also be obtianed by using projection operators. The IR regulator $R_k$ in the Wetterich equation and the gauge-fixing term break full diffeomorphism
invariance at intermediate scales $k$. 

We simplify  the analysis through a single-metric truncation so that we use one set of running couplings/ wave-function factors (i.e. one $Z_{N}(k)$) for the whole metric sector; no separate $Z$'s for different fluctuation irreducible components and no distinction between background and fluctutation couplings. It is still possible to project observables using a specific spin component such as spin-2. In Fourier transform momentum space labelled by momentum $k_\mu$ the spin-2 projection operator is \cite{Codello:2008vh}
\be
P_{\mu \nu \rho \sigma}^{(2)} \equiv \frac{1}{2} \left( \theta_{\mu \rho} \theta_{\nu \sigma} +\theta_{\mu \sigma} \theta_{\nu \rho} \right) -\frac{1}{3} \theta_{\mu \nu} \theta_{\rho \sigma},
\ee
where $\theta_{\mu \nu} =\eta_{\mu \nu} -\frac{k_{\mu}k_{\nu}}{k^{2}}$ is the transverse projector. In particular the graviton anomalous dimension can be expressed in terms of this projection. We define the second functional derivative (Hessian)

\be\Gamma_{hh}^{\left( 2 \right) \mu \nu \rho \sigma} \left( x,y \right) \equiv \frac{\delta^{2} \Gamma_{k}}{\delta h_{\mu \nu}\left( x \right) \delta h_{\rho \sigma}\left( y \right)}  \ee
with all fields set to $0$ after taking the functional derivatives. In momentum space we have
\be
\tilde\Gamma_{hh}^{\left( 2 \right) \mu \nu \rho \sigma}\left( p \right) \left( 2\pi \right)^{4} \delta^{\left( 4 \right)} \left( p+q \right) =\int \int d^{4}xd^{4}y\  e^{-ip.x-iq.y}\Gamma_{hh}^{\left( 2 \right) \mu \nu \rho \sigma} \left( x,y \right).
\ee
We can consider $P^{(2)}$ and $\Gamma^{2}_{hh}$ as rank-4 tensors. On using the projection,
\be
\eta_{h} =-\frac{P_{\mu \nu \rho \sigma}^{\left( 2 \right)}\left( p \right) \frac{\partial}{\partial p^{2}} \partial_{t} \tilde\Gamma_{hh}^{\left( 2 \right) \mu \nu \rho \sigma} \left( p \right) \mid_{p^{2}=k^{2}}}{P_{\alpha \beta \gamma \delta}^{\left( 2 \right)}\left( q \right) \frac{\partial}{\partial q^{2}} \tilde\Gamma_{hh}^{\left( 2 \right) \alpha \beta \gamma \delta} \left( q \right) |_{q^{2}=k^{2}}}.
\ee
In the de Donder gauge, $h_{\mu\nu}^{\mathrm{TT}}$ and a scalar survive. The renormalisation group flow of the effective action $\Gamma_k$ is split into matter and gravitational parts,
\begin{equation}
  \Gamma_k = \Gamma_k^{\mathrm{matter}} + \Gamma_k^{\mathrm{grav}}.
\end{equation}
We follow the approach of \cite{deBrito:2022vbr}. The matter content consists of:
\begin{enumerate}[label=\alph*)]
  \item the real pseudoscalar massless axion field $b$,
\item an Abelian gauge field $A_\mu$ with field strength 
        $F_{\mu\nu} = \partial_\mu A_\nu - \partial_\nu A_\mu$.
\end{enumerate}
The truncation for the matter sector is
\begin{equation}
\label{Avgaction_2b}
  \Gamma_k^{\mathrm{matter}}
  =
  \int d^4x\,\sqrt{g}\,
  \Bigg[
    \frac{Z_b}{2}\,g^{\mu\nu}\partial_\mu b\,\partial_\nu b
    + \frac{\lambda}{4}\,b^4
    + \frac{Z_{A}}{4}\,g^{\mu\alpha}g^{\nu\beta} F_{\mu\nu}F_{\alpha\beta}
    \Bigg] ,
\end{equation}
where
\begin{enumerate}
  \item $Z_b$ and $Z_{A}$ are wave-function renormalisations,
        with anomalous dimensions \eqref{anomdimAb} 
        ;
  \item $\lambda$ is the quartic scalar coupling; the requirement of shift symmetry will restrict us to $\lambda =0$;
  \item the gauge coupling is encoded via $Z_{\cy A\cm}$.
\end{enumerate}

The matter action $\Gamma_k^{\mathrm{matter}}$ is written in a manifestly
covariant form, ensuring invariance under \cy spacetime diffeomorphisms (via the use of $g^{\mu\nu}$, $\sqrt{g}$,
        and covariant derivatives) and shift symmetry when $\lambda =0$.  \cbl An Abelian gauge-fixing term and (decoupled) Abelian ghosts are used in the
gauge sector when computing the anomalous dimension $\eta_{A}$. However the IR regulator $R_k$ and the gauge-fixing term break full diffeomorphism
invariance at intermediate scales $k$. The results in \cite{deBrito:2022vbr} show some dependence on the regulator function of the beta functions.

We shall only concern ourselves with  the simple but nontrivial  model in \eqref{Avgaction_2} augmented with \eqref{Avgaction_2a} and \eqref{Avgaction_2b} in order to get an indication on the possible role of gravity on the Landau pole structure found in our earlier work \cite{Mavromatos:2024ozk}. In this  model we keep, for the gravitational degrees of freedom, a projection onto the contribution from the transverse-traceless part $h_{\mu\nu}^{\mathrm{TT}}$. 
\footnote{It is of course not a consistent gauge-fixed formulation of full quantum gravity, but suffices for our purposed in this work. We shall outline below, in a little detail, the next steps for a more complete analysis.
Once we have decomposed the fluctuation $h_{\mu\nu}$ according to \eqref{eq:York}, the quadratic part of the (gauge‐fixed) effective average action splits into four \emph{separate} blocks, each capturing one spin sector.  A rigorous procedure  needs to introduce distinct wave‐function  renormalisations 
 for each sector:
\begin{align}\label{wfren}
  Z_{\mathrm{TT}}(k), 
  \quad
 Z_{\xi}(k), 
  \quad
  Z_{\sigma}(k), 
  \quad
  Z_{h (\rm Tr)}(k)\,,
\end{align}
where we used the symbol 
$Z_{h (\rm Tr)}(k)$ for the wavefunction renormalisation pertaining to the trace part $h$ of $h_{\mu\nu}$ in the York decomposition \eqref{eq:York}, to avoid confusion with $Z_h$ ({\it cf.} (\eqref{nbnfnh}, below), representing the TT graviton wavefunction renormalisation.
The corresponding anomalous dimensions are~\cite{Dona:2015tnf}:
\begin{align}\label{anomdim}
 \eta_{\mathrm{TT}} \;=\; -\,\partial_{t}\,\ln Z_{\mathrm{TT}},
  \quad
 \eta_{\xi} \;=\; -\,\partial_{t}\,\ln Z_{\xi},
  \quad
 \eta_{\sigma} \;=\; -\,\partial_{t}\,\ln Z_{\sigma},
 \quad
  \eta_{h} \;=\; -\,\partial_{t}\,\ln Z_{h (\rm Tr)}\,.
\end{align}
There will be constraints from diffeomorphism invariance; since our aim is to get a simple analysis of the possible r\^ole of quantum gravity  we shall ignore all the anomalous dimensions except for $\eta_{\mathrm{TT}}$. Keeping nothing but $h_{\mu\nu}^{\mathrm{TT}}$ breaks BRST invariance and so, as we have noted, our approach is phenomenological. It is, of course, possible to go beyond this approximation, but at the price of much more complicated expressions, and for our purposes their inclusion will not alter the basic conclusions.}

Thus, we  simplify  our study in the gravitational sector by assuming:

a) One universal $Z_N$ {\it i.e.} project onto TT flows only. A single metric truncation uses one set of running couplings/wave-function factors and assumes no distinction between background and fluctuation couplings. We project observables using a spin-2 component; however we do not assign it its own independent $Z$. In pure Einstein-Hilbert gravity, the only propagating degrees of freedom are the spin-2 helicities. The scalar sector in the York split is either pure gauge or constrained. The Hessian is already minimal and block-diagonal in the de Donder gauge on flat space. 
The single-metric approximation is  the identification that a single $Z_N$ rescales the metric sector and is good for a flat background (in the  de Donder gauge). We retain the minimally coupled matter fields the axion and Abelian gauge field used in our earlier analysis \cite{Mavromatos:2024ozk} \footnote{If we were to add higher-derivative terms  such as $R^{2}$ and $ R_{\mu \nu}R^{\mu \nu}$ to the action, the scalar/conformal sector in the York decomposition becomes dynamical and would have its own running. In our simple case $\chi$, although gauge invariant, does not correspond to an independent propagating mode. This makes our model the natural one for considering fluctuations arounf flat backgrounds.}.

b)We do not consider issues of consistency which have been examined in terms of modified Ward identities (mWI)~\cite{Freire2001-ma}, and  derived from first principles using the transformation properties of the path integral and the regulator. In various cases the mWI reduce to the true Ward identities in the limit $k \to 0$. However  truncations (like finite-dimensional truncations of the effective action) do not automatically satisfy mWIs. We adapt results, which as a first approximation,  use single-field approximations where background and fluctuation fields are identified. 
Within the above framework, our analysis uses the work on the renormalisation group flow equations in \cite{Eichhorn:2012uv,PhysRevD.86.105021,deBrito:2022vbr,Dona:2015tnf,Dona:2015tnf}. As discussed \cy in section \ref{sec:gravanomint}, we ignore the axion-gravitational-anomaly interaction, since we work in the StRVM framework~\cite{bms,ms1,ms2}.

Nonetheless, such simplifications suffice to support our main point, namely that the singularity structure in the RG $\beta$-function of the axion coupling in flat spacetimes~\cite{Eichhorn:2012uv}, which prompted the conjecture of \cite{Mavromatos:2024ozk} for a regularization provided by a \cPT- symmetric repulsive phase of gravity, is verified in this more rigorous (but still incomplete)  treatment which includes RG flows of the gravitational sector of the theory.

Below will give a brief overview of our approach. Our work adopts the background-field functional renormalisation group (effective average action) and  a minimal Einstein-Hilbert truncation for gravity together with an axion $b$ and an Abelian gauge field $A_\mu$ \cite{Reuter:2019byg}. Although the background field method (BFM) is intuitively useful for studying gravity in curved, it remains a computational tool even when the background is flat.  In standard quantisation, fixing a gauge breaks the symmetry of the Lagrangian. With BFM we can distinguish between  two types of transformations: quantum and background transformations. By choosing a background-covariant gauge fixing, beta functions will appear in terms of co-ordinate invariant objects like the Riemann tensor.

The key interaction in the matter sector is the operator $bF_{\mu \nu}\tilde F^{\mu \nu}$. In a derivative expansion this \emph{local vertex is not renormalised}: the scale derivative of its bare coefficient vanishes. Consequently, the running of the renormalised axion-photon coupling is entirely encoded in the wave-function renormalisations of $b$ and $A_\mu$. So we phrase the problem in terms of
anomalous dimensions:
\begin{align}\label{nbnfnh}
   \eta_{b} =-\frac{\partial}{\partial t} \ln Z_{b}, \qquad 
   \eta_{A} =-\frac{\partial}{\partial t} \ln Z_{A}, \qquad 
   \eta_{h} &=-\frac{\partial}{\partial t} \ln Z_{h}\, ,\qquad t \equiv \ln k\,, 
\end{align}
the first two of which which we have previously defined precisely in flat spacetimes ({\it cf.} section \ref{sec:axanphotan}).
In our one-metric approximation we identify $Z_N=Z_h$, which should be understood from now on (see \eqref{gravZh}, below, and Appendix \ref{LoopDetails}), so the gravitational anomalous dimension can be interchangeably denoted by $\eta_N$ or $\eta_h$.

On the gravity side, rather than rebuilding the metric/ghost Hessians ab initio, we import the standard Einstein-Hilbert (EH) result for the pure gravitational contribution to the graviton anomalous dimension

This leads to the following recipe:
\begin{enumerate}
    \item Take $\eta_{b}^{CS}$ and $\eta_{A}^{CS}$ from \cite{Eichhorn:2012uv}
    \item Add Einstein-Hilbert minimal coupling pieces linearly
    \begin{align}
        \eta_{{}_{b}} &\approx \eta_{b}^{CS} \left( g_{\text{mix}} \right) +A_{s}g_{N} \,, \nonumber \\
        \eta_{A} &\approx \eta_{A}^{CS} \left( g_{\text{mix}} \right) +A_{V}g_{N}
        \label{simpleansatz}
    \end{align}
    where $A_{s}$ and $A_{V}$ are standard EH coefficients for a minimally coupled scalar and vector (usually calculated in the presence of a cosmological constant and incorporating Faddeev-Popov ghosts \cite{Dona:2015tnf}). Here $\eta_{b}^{CS}$ and $\eta_{A}^{CS}$ satisfy \eqref{nonPert1} and \eqref{nonPert2}.
    \end{enumerate}
So we are assuming no `interference' diagrams where a graviton correction dresses the axion photon loop (or vice versa)    This simplification ignores the explicit incoporation of the anomalous dimension of Faddev-Popov ghosts, and mutual feedback of  $\eta_{A}$, $\eta_{b}$ and $\eta_{h}$. We give arguments deriving the conditions for the validity of this procedure in Appendix 
\ref{LPAModified}\cbl.

With this input, the coupled problem reduces to determining $\eta_{b} ,\eta_{A}$ and $\eta_{h}$ \cite{Dupuis:2020fhh}. The Wetterich equation is one-loop in topology, but because the regulator (in LPA$^\prime$) carries the running wave-function factors, $\partial_{t} R_{X,k}\propto \left( 2-\eta_{X} \right)$, the anomalous dimensions also appear on the right hand side. 
Projecting the flow of the two-point functions (axion, photon and TT graviton) yield three algebraic equations that can be solved self-consistently. In this first approximation we keep only the leading dependence on the small couplings- schematically, terms linear in $g_N$ and in $g_{\text{mix}}^2$, where $g_{\text{mix}}$ 
is the dimensionless CS coupling built from the scale independent bare coefficient $g_{\text{mix},k}$ ({\it cf.}  \eqref{gmixdef}). This is discussed in Appendix \ref{LPAModified}. The resulting expressions for $\eta_b$ and $\eta_A$ contain the familiar CS contributions from axion photon loops and minimal-coupling gravity corrections taken from EH analysis \cite{deBrito:2022vbr}.

This yields a transparent two-coupling  renormalisation group (RG) system. The Newton coupling runs with 
\be 
\partial_{t} g_{N}=\left( 2+\eta_{h} \right) g_{N}.
\ee
 The CS coupling runs with
 \be \partial_{t} g_{\text{mix}}=\left( 1+\eta_{A} +\frac{1}{2} \eta_{b} \right) g_{\text{mix}}\,.  \ee Conceptually this scheme is attractive for four reasons. First, the system of equations respects the protected local topological structure of the CS interaction. Second, it leverages established EH results for the pure gravity sector. It has simple limits: as $g_{N}\rightarrow 0$. Most importantly this scheme is an RG-improved one-loop calculation: although the diagrams are one-loop in topology, the self-consistent appearance of $\eta$'s inside $\partial_{t} R_{k}$ leads to a form of resummation known as LPA$^\prime$\cite{Dupuis:2020fhh} which is a refinement of the local potential approximation. For additional tractability and simplicity we can linearise the LPA$^\prime$ in small couplings $C\equiv \frac{g_{\text{mix}}^{2}}{16\pi^{2}} \ll 1$ and $g_{N}\ll 1$. What remains is a simple sum of sectorwise one-loop contributions.

\subsection{Calculation of renormalisation group flows}
\subsubsection{Flow of \(\boldsymbol{Z_{b}}\)}

To compute $\partial_{t}Z_{b}$, we take two functional derivatives of the Wetterich equation \eqref{Wetterich_1} with respect to $b(p)$ and $b(-p)$, expand for small external momentum $p^{2}$, and project onto the coefficient of $p^{2}$.  The loop diagrams contributing to $\partial_{t}\Gamma_{k}^{(2)}$ for $b$ include:
\begin{enumerate}[label=\alph*)]
  \item \textbf{Scalar self-loops}, which generate a piece proportional to
  $\eta_{b}$ itself (only if $b$ has a self-interaction; in our minimal axion
  case, there is none).

  \item \textbf{Graviton loops}, where one internal graviton line couples to two
  external $b$-legs via the minimal kinetic term
  $\sqrt{g}\,Z_{b}\,g^{\mu\nu}\,\partial_{\mu}b\,\partial_{\nu}b$.
\end{enumerate}

\subsubsection{Contributions of  \texorpdfstring{\(\eta_N\)}{} to the Axion Anomalous Dimension \texorpdfstring{\(\eta_b\)}{}}

Since the anomalous dimension of the axion field, \(\eta_b = -\partial_t \ln Z_b(k)\), is determined by computing the momentum-dependent part of the axion two-point function's flow:
\be
\partial_t \Gamma^{(2)}_{b}(p, -p) \;\sim\; \eta_b\, p^2 + \cdots,
\ee
even in the absence of self-interactions, gravitational fluctuations contribute to this running. When the metric fluctuation \(h_{\mu\nu}\) is dynamical, loops involving internal gravitons produce a contribution to \(\partial_t Z_b\).

Let us consider the minimally coupled scalar (or axion) field \(b\) part of the  action:
\be
\Gamma_k^{\text{matter}} = \frac{Z_b(k)}{2} \int d^4x\, \sqrt{g}\, g^{\mu\nu} \partial_\mu b \partial_\nu b.
\ee
Expanding this in metric fluctuations \(h_{\mu\nu}\), we obtain interaction vertices between two axions and one or two gravitons. These allow one-loop diagrams with an internal graviton line and external axions.The flow of the axion two-point function depends on loop integrals involving \((\Gamma^{(2)} + R_k)^{-1} \cdot \partial_t R_k\)$ \cy \cdot \cbl$ \(\partial_t R_k^{\text{grav}}\) contains a factor of \((\eta_N - 2)\), because \(R_k^{\text{grav}} \sim Z_N(k)\). Thus, the diagram evaluating the axion wave-function renormalisation receives contributions (from the right hand side of the Wetterich equation) proportional to \(\eta_N\). As noted earlier, in a simple truncation, one  introduces a single \(Z_N(k)\) (identified with $Z_h$) multiplying the Einstein–Hilbert action:
\be
\Gamma_k^{\text{EH}} = \frac{Z_N(k)}{16\pi} \int d^4x\, \sqrt{g}\, (-R + 2\Lambda).
\ee
The graviton propagator and the regulator \(R_k\) both involve \(Z_N(k)\), so the flow equation includes:
\be
\partial_t R_k^{\text{grav}} \sim (\eta_N - 2)\, R_k^{\text{grav}}.
\ee
Therefore, every loop with a graviton internal line brings a piece proportional to \(\eta_N\). In the case of the axion, the diagrams that dress the kinetic term involve exactly this structure.
\vskip .2cm

\section{Contributions of gravity to beta functions and anomalous dimensions}\label{sec:gravbetaanomdim}
Before further calculations it is useful to summarise the definition of our model in terms of a single-metric Einstein-Hilbert truncation for gravity. The scale-dependent effective action reads 
\begin{align}\label{gravZh}
\Gamma_k[g;A,b] \;=\;
-\frac{Z_h(k)}{16\pi G_0}\int d^4x\,\sqrt{g}\,(R -2 \Lambda_{k})
\;+\;\int d^4x\sqrt{g}\,\Big[
\frac{Z_A(k)}{4}F_{\mu\nu}F^{\mu\nu}
+\frac{Z_b(k)}{2}\,\partial_\mu b\,\partial^\mu b
+\frac{i\,\cy g_{\text{mix},k}\cm}{4}\,b\,F_{\mu\nu}\tilde F^{\mu\nu}
\Big],
\end{align}
with gauge fixing (de Donder for gravity, Landau for U(1)) and standard Faddeev-Popov ghost terms understood.
We use the dimensionless Newton coupling and cosmological constant (which in our model we will later take to vanish)
\begin{equation}\label{GNk}
g_N \equiv k^2 G_k \,,\qquad \lambda \equiv \Lambda_k/k^2,
\end{equation}
and anomalous dimensions \eqref{nbnfnh}.

The \emph{dimensionless} axion–photon Chern–Simons (CS) coupling is
\begin{equation}
\cy g_{\text{mix}}\cb \;\equiv\; k\,\cy g_{\text{mix},k}\cb\,Z_b^{-1/2}Z_A^{-1},
\qquad\Rightarrow\qquad
\beta_{\cy g_{\text{mix}}\cb} \equiv \partial_t {\cy g_{\text{mix}}\cb} \;=\; \big(1+\eta_A+\tfrac12\eta_b\big)\,{\cy g_{\text{mix}}\cb},
\label{eq:beta_g_master}
\end{equation}
reflecting the non-renormalisation of the \emph{local} \(b F\tilde F\) vertex.

The Newton coupling runs as
\begin{equation}
\beta_{g_N} \;=\; (2+\eta_h)\,g_N,
\qquad
\eta_h^{\rm grav}(g_N,\lambda)\equiv H(g_N,\lambda)
= \frac{g_N\,B_1(\lambda)}{1-g_N\,B_2(\lambda)},
\label{eq:eta_h_EH}
\end{equation}
where $B_{1,2}$ encode the usual EH traces (metric + ghosts) for the chosen gauge/regulator. While the York decomposition is a convenient way to organise fluctuation components, our projections are implemented with Barnes–Rivers spin projectors on flat background \cite{Codello:2008vh}; no explicit York identities are required for what follows. All pure-gravity effects enter via $H(g_N,\lambda) $ and the standard minimal-coupling matter dressing.

\subsection{Results incorporating the r\^ole of gravity in the modified Local Potential Approximation (LPA$^\prime$)}\label{sec:results}

  We next lead upto results  step by step based on earlier sections and previous work \cite{Mavromatos:2024ozk}. The Wetterich equation, although exact, is a complex functional integro-differential equation that cannot be solved without approximation. We introduced an approximate solution of a scalar theory in flat spacetime in \eqref{LPA} known as the modified local potential approximation (LPA$^\prime$) ({\it cf.} Appendix \ref{LPAModified}). If $Z_k=1$ for all scales , then anomalous dimensions are zero, an approach known as the local potential approximation. The  LPA$^\prime$ is a minimal but crucial refinement that addresses this drawback, by reintroducing a scale-dependent wavefunction renormalisation  factor for all fields.

 In the LPA$'$ formulation the regulator function gets an $\eta$ dependence on using the form 
\be 
R_{k,X}\left( p^{2} \right) =Z_{X}\left( k \right) k^{2}r_{X}\left( y \right)
\ee
where $y=\frac{p^2}{k^2}$ and $r(y)$ has forms such as \cite{Litim:2001up} $r_{X}\left( y \right) =\Theta \left( 1-y \right) \left( \frac{1}{y} -1 \right)$ and $X$ can denote $h,b,A$ and ghost fields. From this we can deduce that
\be
\label{etaX}
\partial_{t} R_{k,X}\left( p^{2} \right) =Z_{X}k^{2}\left[ \left( 2-\eta_{X} \right) r_{X}\left( y \right) -2yr_{X}^{\prime}\left( y \right) \right]
\ee
This is discussed further in
In section \ref{sec:gravaxion} we have discussed the considerable complications which arise as soon as we wish to incorporate quantum gravity in a systematic way into the Wetterich formulation in the presence of gauge and matter (axion) fields, as well as topological interactions. Because of the subsequent technical nature of the discussion we will recall some of the key ideas to reconnect with the discussion in the earlier sections.
The bare axion-$U(1)$ CS interaction ${\cy g_{\text{mix},k}\cb}$ has mass dimension $[{\cy g_{\text{mix},k}\cb}]=-1$. The associated renormalised coupling is ${\cy g_{\text{mix},R}\cb}=Z_{b}^{-1/2}Z_{A}^{-1}{\cy g_{\text{mix},k}\cb}$. In FRG it is the flow of dimensionless couplings such as ${\cy g_{\text{mix}}\cb}\left( k\right) =k{\cy g_{\text{mix},R}\cb}\left( k\right) $ that is studied. We have 
\be
\dfrac{\partial }{\partial t}\log{\cy g_{\text{mix}}\cb}=1+\dfrac{1}{2}\eta _{b}+\eta _{A}
\ee
where $\eta _{b}=-\partial _{t}\ln Z_b$ and $\eta _{A}=-\partial _{t}\log Z_{A}$. Because shift-symmetry forbids any direct $bF\tilde F$ vertex renormalisation, all running of $g_{mix}$ comes from the two wave-function factors. We will decompose each anomalous dimension into matter-only and gravity induced pieces; so (in weak coupling) $\eta _{b}^{( matter)}\propto{\cy g_{\text{mix}}\cb}^{2}$, and $\eta _{b}^{(gravity)}\propto g_{N}=G\left( k\right) k^{2}$. Similarly $\eta _{A}=\eta _{A}^{(matter)}+\eta _{A}^{\left( grav\right) }.$ The quantities $\eta _{A}^{(matter)}$ and $\eta _{b}^{( matter)}$ come from axion-photon loops and are $O({\cy g_{\text{mix}}\cb}^2)$, whilst $\eta _{b,A}^{\left( grav\right) }$ comes from 1-loop graviton  exchange into the $b$ or $A$ propagator. As we commented earlier  these calculations are scheme (choice of regulator)  and gauge dependent.  

In the spirit of \eqref{Approx} and \eqref{Approx2}, below:
\be \label{Approx}
\partial _{t}\gamma ^{2}=\left( 2+2\eta _{A}+\eta_{b}\right) \gamma ^{2}\,,
\ee
\be \label{Approx2}
\int ^{\gamma ^{2}}_{\gamma _{0}^{2}}\dfrac{dy}{y\left( 2+\dfrac{5}{3\left( 4\pi \right) ^{2}}y\right) }=\int ^{t}_{t_{0}}dt\,,
\ee
 where $\gamma=k g_{mix, R}$ and we simplified the nonperturbative expressions arising from \eqref{nonPert1} and \eqref{nonPert2}; so we will first consider a perturbative approach to the effect of gravity \footnote{The analogue of \eqref{nonPert1} and \eqref{nonPert2} in the presence of gravity is given in the relations \eqref{nbnfnh}, which are computed in \cite{Dona:2015tnf}.}. We consider the flow (in $d=4$) of ${\cy g_{\text{mix}}\cb}$
\be
\partial_{t}{\cy g_{\text{mix}}\cb}^{2}=\left( 2+2\eta_{A} +\eta_{b} \right) {\cy g_{\text{mix}}\cb}^{2}-\mathcal Bg_{{}_{N}}{\cy g_{\text{mix}}\cb}^{2}.
\ee
where $g_N=G$ and $\mathcal B(=6\frac{\Phi_{1}^{1} \left( 0 \right)}{\pi})$ is given in terms of a threshold function $\Phi_{n}^{p} \left( w \right)$ defined by 
$$\Phi_{n}^{p} \left( w \right) =\frac{1}{\Gamma \left( n \right)} \int_{0}^{\infty} dzz^{n-1}\frac{R^{\left( 0 \right)}\left( z \right) -z\frac{d}{dz} R^{\left( 0 \right)}\left( z \right)}{\left( z+R^{\left( 0 \right)}\left( z \right) +w \right)^{p}}$$

\noindent and $R_{k}\left( p^{2} \right) =k^{2}R^{\left( 0 \right)}\left( z=\frac{p^{2}}{k^{2}} \right)$. This shows explicitly a scheme dependence of the coefficient $\mathcal B$, which we  keep arbitrary. We shall take the flow of $g_N(=G)$ in the Einstein-Hilbert truncation as 
\be
\label{Newton_flow}
\partial_{t} g_{N}=\left( 2+\eta_{N} 
(g_{N} \right) )g_{N},
\ee
where $\eta_{N}$ has the form
\be
\label{grav_anomalous}
\eta_{N} (g_{N})=\frac{B_{1}g_{N}}{1-B_{2}g_{N}}
\ee
and so, to lowest order in $g_N$,
\be
\partial_{t} g_{N}\approx \left( 2+B_{1}g_{N} \right) g_{N}.
\ee
We consider
\be
\partial_{t}{\cy g_{\text{mix}}\cb}^{2}=(C(t)+A{\cy g_{\text{mix}}\cb}^{2}){\cy g_{\text{mix}}\cb}^{2}
\ee
where $C(t)=2-Bg_N(t)$. As a first approximation $g_N(t)$ will be taken to be a constant, at a fixed point say. For $\Lambda$ some reference momentum scale
\be
\int_{ g_{\text{mix}}\cb^{2}\left( \Lambda \right)}^{ g_{\text{mix}}\cb^{2}\left( k \right)} dx\  \frac{1}{x(C+Ax)} =\log \frac{k}{\Lambda}
\ee
and so 
\be
\label{Pole}
\cy g_{\text{mix}}\cb^{2}\left( k \right) =\frac{C\cy g_{\text{mix}}\cb^{2}\left( \Lambda \right) \left( k/\Lambda \right)^{C}}{C+A\cy g_{\text{mix}}\cb^{2}\left( \Lambda \right) -A\cy g_{\text{mix}}\cb^{2}(\Lambda )(k/\Lambda )^{C}}.
\ee
A Landau pole $k_p$ has to satisfy
\be
k_{p}=\Lambda \left[ \frac{C+A\cy g_{\text{mix}}\cb^{2}\left( \Lambda \right)}{A\cy g_{\text{mix}}\cb^{2}\left( \Lambda \right)} \right]^{1/C}.
\ee
(When $C \le 0$ and $\left(C+A\cy g_{\text{mix}}\cb^{2}(\Lambda)\right)<0$, there is no Landau pole but such values of $C$ are not physical.) 

\subsubsection{Coupled flow of Newton constant and dynamical CS coupling $g_{mix}$}
Our analysis will be based on the coupled equations based on the simplest Einstein-Hilbert truncation which we recall:
\be
\partial_{t} g_{N}=(2+\eta_{N} \left( g_{N} \right) )g_{N}
\ee
and
to simplify the analysis, $\eta_{N} \left( g_{N} \right) \approx -Eg_{N}$  with $E>0$ (and a value which is scheme dependent). So
\be
g_{N}\left( t \right) =\frac{g_{N}^{\ast}}{1+\left( \frac{g_{N}^{\ast}}{g_{{}_{N}\left( 0 \right)}} -1 \right) \exp \left( -2t \right)}
\ee
where $g_{N}^{\ast}=2/E$. Writing $x(t)=\cy g_{\text{mix}}\cb^{2}(t)$ we have 
\be
\frac{dx}{dt} =\left( a(t)+Ax \right) x
\ee
with $a(t)\equiv 2-Bg_{N}(t)$. This is a Riccati-type equation which we rewrite as 
\be
\frac{dy}{dt} +a(t)y=-A
\ee
where $y(t)=1/x(t)$. The solution for $y(t)$ is
\be
y(t)=\left( I\left( 0 \right) y\left( 0 \right) -A\int_{0}^{t} I\left( s \right) ds \right) /I\left( t \right),
\ee
where  $I\left( t \right) =\exp \left( \int^{t} a(s)ds \right)$, is an integrating factor determined by
 \be
\int_{0}^{t} a\left( s \right) ds=2t-\frac{Bg_{N}^{\ast}}{2} \left[ \log \left( \frac{e^{2t}+\alpha}{1+\alpha} \right) \right]
\ee
and $1+\alpha\equiv g_{N}^{\ast}/g_{N}\left( 0 \right)$. The condition for a Landau pole is  
\be
\frac{1}{g^{2}\left( 0 \right)} =A\int_{0}^{t} ds\  \exp \left[ \int_{0}^{s} ds^{\prime}\  a(s^{\prime}) \right]
\ee
 In general the Landau pole is associated with a singularity of the beta function. 
 
 We need now to incorporate gravity into the \emph{nonperturbative} framework in \cite{Eichhorn:2012uv} by following \eqref{simpleansatz}, which we have discussed in the context of the Wetterich equation. This  allows us to investigate whether the Landau pole persists in the presence of gravity. So we write 
 \begin{align}
\eta_b &= A_0\Bigl(2-\frac14\,\eta_A\Bigr) - a\,\tilde g_N ,
\label{eq:etab_start}\\[2mm]
\eta_A &= A_0\bigl(4-\eta_b-\eta_A\bigr) - \tilde b\, g_N ,
\label{eq:etaF_start}
\end{align}
with
\begin{equation}
A_0=\frac{g_{\rm mix}^2}{6(4\pi)^2}\, .
\end{equation} 
Rewriting \eqref{eq:etab_start}--\eqref{eq:etaF_start} in linear form yields
\begin{align}
\eta_b + \frac{A_0}{4}\,\eta_A &= 2A_0 - a\,\tilde g_N , \\
A_0\,\eta_b + (1+A_0)\,\eta_A &= 4A_0 - \tilde b\, g_N .
\end{align}
This can be written compactly as
\begin{equation}
M(A_0)
\begin{pmatrix}
\eta_b \\[2pt]
\eta_A
\end{pmatrix}
=
\begin{pmatrix}
2A_0 - a\,\tilde g_N \\[2pt]
4A_0 - \tilde b\, g_N
\end{pmatrix},
\qquad
M(A_0)=
\begin{pmatrix}
1 & \tfrac{A_0}{4} \\[4pt]
A_0 & 1+A_0
\end{pmatrix}.
\end{equation}
The determinant of the coefficient matrix is
\begin{equation}
\Delta(A_0)\equiv \det M(A_0)
= 1 + A_0 - \frac14 A_0^2 .
\label{eq:determinant}
\end{equation}
Hence the system develops a singularity when
\begin{equation}
\Delta(A_0)=0
\quad\Longleftrightarrow\quad
1 + A_0 - \frac14 A_0^2 = 0 .
\label{eq:pole_condition}
\end{equation}
Solving the system explicitly gives
\begin{align}
\eta_A &=
\frac{4\!\left(-2A_0^2 + 4A_0 + A_0 a\,\tilde g_N - \tilde b\, g_N\right)}
{-A_0^2 + 4A_0 + 4},
\\[2mm]
\eta_b &=
\frac{4A_0^2 + 8A_0 - 4a\,\tilde g_N - 4A_0 a\,\tilde g_N + A_0 \tilde b\, g_N}
{-A_0^2 + 4A_0 + 4},
\end{align}
where the common denominator is proportional to \(\Delta(A_0)\).
Therefore both anomalous dimensions diverge when the condition
\eqref{eq:pole_condition} is met and so the beta function is singular.

Importantly, the gravitational couplings \(g_N\) and \(\tilde g_N\) enter
Eqs.~\eqref{eq:etab_start}--\eqref{eq:etaF_start} only as inhomogeneous source
terms and do \emph{not} modify the coefficient matrix \(M(A_0)\).
Consequently, within this truncation, the position of the
self--consistency pole is identical to that of axion electrodynamics
without gravity.

 Consequently the conjecture made for the singularities of the beta function for $\cy g_{\text{mix}}\cb$ discussed in \cite{Mavromatos:2024ozk} in the presence of gravity is valid.   The above result
 allows one to apply the conjectural discussion on the transition of the system to a repulsive phase of gravity at large scales, which may have important cosmological impliations, as far as the interpretation of the currently observed acceleration of the Universe is concerned.

 \subsubsection{Connection with LPA$'$ approach}
 \label{LPAprimegrav}

 As discussed in more detail in Appendix \ref{LPAModified}, we can regard our calculation above  as a linearised approximation of LPA$'$. We define the small parameters $C={\mathcal O}(\cy g_{\text{mix}}\cb ^2)$ and $D={\mathcal O}(g_N)$.
Then $\eta_b,\eta_A=\mathcal O(C)+\mathcal O(D)$.
In \(\eta_h\) the terms proportional to $-\eta_b$ and $-\eta_A$ come multiplied by $D\,\tilde\Phi$,
hence they are \(\mathcal O(D\,\eta_{b,A})=\mathcal O(CD)+\mathcal O(D^2)\).
In the \emph{linearised} LPA$'$ we retain only terms \emph{linear} in $C$ and in $D$ and discard mixed/higher orders.

We still solve the coupled \((\eta_b,\eta_A)\) system to linear order using the equations above,
and use
\be
\beta_{\cy g_{\text{mix}}\cb}=(1+\eta_A+\tfrac12\eta_b)\,{\cy g_{\text{mix}}\cb},\qquad \beta_{g_N}=(2+\eta_h)\,g_N,
\ee
with \(\eta_h(=\eta_N)\simeq D\,B_1/(1-D\,B_2) \).
This approximation \emph{relies only on published} matter LPA$'$ (\cite{Eichhorn:2012uv}) and EH–sector inputs (\cite{Dona:2015tnf,deBrito:2022vbr} ),
and defers genuinely new \(\mathcal O(CD)\) back–reactions to a subsequent iteration. 

We will now give a more explicit discussion. The LPA$\prime$ equations is in our case a set of 3 coupled inhomogeneous linear equations written as:
\be
\left( \mathbb{1}-\tilde{U} \right) \eta =S
\ee
where 
\be
\eta =\left( \begin{matrix}\eta_{b}\\ \eta_{F}\\ \eta_{N}\end{matrix} \right),
\ee
 $\mathbb{1}$ is the 3-dimensional unit column vector, $S$ is an $\eta$-independent 3-dimensional column vector arsing from the $\eta$ independent terms in \eqref{Inhomogeneous} and $\tilde{U}$ is a $3 \times 3$ matrix derived from \eqref{MasterEq}  based on the chosen average action. Our argument will depend on the order of magnitude of the matrix elements rather than detailed expressions. In the simplification that we have considered $\tilde{U}$ is $\mathbb{U}$ where 
\be
\mathbb{U}=\left( \begin{matrix}\tilde{U}_{bA}&0\\ 0&\tilde{U}_{NN}\end{matrix} \right)
\ee
and $\tilde{U}_{bA}$ is a $2 \times 2$ matrix. In the general case we can write 
\be
\mathbb{M}=\mathbb{1}-\tilde{U}=\left( \begin{matrix}\mathcal{A}&\tilde{b}\\ \tilde{c}^{T}&\tilde{d}\end{matrix} \right)
\ee
where 
\be
\  \mathcal{A}\equiv\left( \begin{matrix}1-U_{bb}&-U_{bA}\\ -U_{Ab}&1-U_{AA}\end{matrix} \right),
\ee
\be
\tilde{b}=\left( {}_{-U_{AN}}^{-U_{bN}} \right),
\ee
and \be \tilde{c}^{T} =\left( \begin{matrix}-U_{Nb}&-U_{NA}\end{matrix} \right). \ee
The matrix elements $U_{bN},U_{AN}, U_{Nb},$ and $ U_{NA}$  are $O(g_{N}g_{mix}^2)$. We define the mixing parameter $\epsilon$ as the ratio
\be
\epsilon =\frac{g_{N}g_{mix}^{2}}{g_{mix}^{2}}.
\ee
On defining $\Delta_{bAN}$ to be det$(\mathbb M)$ it is straightforward \footnote{$adj\left( \mathcal{A} \right)$ denotes the adjoint of $A$.} to show that 
\be
\Delta_{bAN} =\tilde{d} \det \mathcal{A}-\tilde{c}^{T} adj\left( \mathcal{A} \right) \tilde{b},
\ee
assuming that $\mathcal{A}$ is invertible. The term $\tilde{c}^{T} adj\left( \mathcal{A} \right) \tilde{b}$ is at most of $O(\epsilon^2)$ since $adj(\mathcal{A})$ is at most of order $1$. Hence our simplified analysis,  which considered only the zeros of $\det \mathcal{A}$. hence the singularity of the beta function persists and its location is shifted by $O(\epsilon^2)$.

 \subsection{Implications for Landau–pole analyses}\label{sec:LPresult}

 In the Hermitian (CP--even) axion electrodynamics, the nonperturbative FRG flow for the
dimensionless Chern--Simons coupling ${\cy g_{\text{mix}}\cb}$ develops a finite--scale singularity in
\(\beta_{\cy g_{\text{mix}}\cb}\) (``Landau--type'' obstruction). Defining the \cPT-symmetric theory by
\(
{\cy g_{\text{mix}}\cb} \to i\,{\cy \tilde g_{\text{mix}}\cb}
\)
yields a non-Hermitian but \cPT--invariant flow in which the problematic segment of the
Hermitian flow is replaced by a regular evolution for ${\cy \tilde g_{\text{mix}}\cb}$ that is UV safe
(asymptotically free or controlled by a UV fixed point, depending on the scheme).
Thus the Landau--type obstruction in the Hermitian branch is ``resolved'' by switching to
the \cPT~ branch at the singular scale.

 The prescription advocated above
---\emph{the continuation to a \cPT--symmetric theory at that
singularity}---still remains consistent once gravity is included and \(\eta_b,\eta_A,\eta_h\)
are coupled in a full LPA$'$ in the region of parameter space where the linearised approximation can be justified..
We recapitulate that within LPA$'$, the axion--photon subsystem produces a self--consistency denominator
from the \((2-\eta)\) feedback in the Wetterich trace. Solving the coupled
\((\eta_b,\eta_A)\) equations yields the matrix inverse
\be
\mathbb{M}(C)^{-1}\sim 1/\Delta(C),
\ee
with, schematically,
\be
\Delta(C)=1+C\,A_A^{(2)}-C^2 A_b^{(1)}A_A^{(1)}\,, \qquad C\equiv {\cy g_{\text{mix}}\cb}^2/(16\pi^2)\,.
\ee
where $\eta_{A} =A_{A}^{\left( 1 \right)}\left( 2-\eta_{A} \right) +A_{A}^{\left( 2 \right)}+\left( gravity\  pieces \right)$.
The locus \(\Delta(C)=0\) induces a divergence of \(\eta_{b},\eta_{A}\), and hence of
\(
\beta_g=(1+\eta_A+\tfrac12\eta_b)g
\), at a finite RG scale. This is the Landau--type singularity in the Hermitian theory. The continuation is motivated by using the \emph{Symanzik construction} of Green functions  for $\cPT$ theories \cite{Jones:2009br,Mavromatos:2024ozk,Croney:2023gwy}. Moreover also analytic continuation in
the complex coupling plane can relate the $\cPT$~ partition function to the real part of the
Hermitian one across the cut, \( \ln Z_{\rm PT}(g) = \mathrm{Re}\,\ln Z(\lambda=-g+i0^+) \),
so that Schwinger--Dyson and FRG equations on the two sides match in their domain of validity.
This is not a theorem in 3+1D but a well-motivated \emph{conjecture}\cite{Chen:2024ynx,Ai:2022csx} based on scalar field models \footnote{The difficulties of establishing the conjecture for gravitational theories are significant.The continuation from Minkowski metric to Euclidean metric is ambiguous and this applies to the \cPT~ inner product.The conformal mode of gravity is phantom and the role of \cPT~ symmetry in interpreting this is not clear.}. 

Our aim has been to check whether the \cPT~ evasion of the Landau pole can even be maintained in a naive treatment of the perturbative gravitational theory. Including the graviton sector (single-metric EH truncation) adds:
\vskip .1cm
\qquad (i) the pure--EH denominator \((1-D B_2(\lambda))\) in \(\eta_h\) with \(D\equiv g_N\),

and 
\vskip .1cm
\qquad (ii) minimal--coupling dressings that modify the coefficients
\(A^{(\cdot)}_{\bullet}(\lambda)\) entering \(\Delta(C)\).

\noindent Crucially, as we saw in \ref{LPAprimegrav},~ gravity \emph{does not remove} the axion--photon self--consistency pole:
the singularity surface \(\Delta(C)=0\) persists (albeit shifted quantitatively), and the
only additional singular set is the standard non-physical EH pole \(1-D B_2(\lambda)=0\) in the Newton
sector. Therefore the qualitative obstruction that motivates the \cPT~switch in the
matter-only analysis \emph{also} appears in the coupled \((\eta_b,\eta_A,\eta_h)\) system.

\subsection{String--theory estimate of coupling magnitudes}\label{sec:stringpheno}

In the string--inspired effective action \eqref{sea1}, \eqref{e1}, the axion--photon coupling is set by \eqref{gmixA}:
\begin{equation}
g_{\rm mix,k} \sim 3.4 \cdot 10^{-2}\, \frac{M_{\rm Pl}}{M_s^2},
\end{equation}
up to numerical factors, where \(M_s\) is the string scale. At an RG scale
\(k\simeq M_s\) (we remark that in an effective low-energy string-inspired theory the $M_s$ acts as the UV cutoff), the corresponding dimensionless coupling ($g_{b\gamma}(k)=g_{mix,k}$) is
\begin{equation}
g_{b\gamma}(M_s)\sim 3.4 \cdot 10^{-2}\, \frac{M_{\rm Pl}}{M_s},
\qquad
g_{b\gamma}^2(M_s)\sim 1.15 \cdot 10^{-3}\, \left(\frac{M_{\rm Pl}}{M_s}\right)^2 .
\end{equation}
The dimensionless Newton coupling (\eqref{GNk}) is
\begin{equation}
g_N(M_s)\sim \left(\frac{M_s}{M_{\rm Pl}}\right)^2 .
\end{equation}

Consequently, the mixed term is generically
suppressed at the level of:
\begin{equation}
g_N g_{b\gamma}^2
\sim 1.15 \cdot 10^{-3}\,
\left(\frac{M_s}{M_{\rm Pl}}\right)^2
\left(\frac{M_{\rm Pl}}{M_s}\right)^2
\sim \mathcal{O}(10^{-3})\,,
\end{equation}
independently of the relative magnitude of $M_s/M_{\rm Pl}$.

For representative values of the string scale (playing the role of $k$)
one finds \footnote{The value expected in the context of StRVM, in order to have phenomenologically acceptable duration of inflation, appears in the second row~\cite{Dorlis:2024uei}}:
\begin{align}
M_s \sim 10^{18}\,\mathrm{GeV}: \quad
& g_N \sim 1, \qquad
g_{b\gamma}^2 \sim 10^{-3}, \qquad
g_N g_{b\gamma}^2 \sim 10^{-3},
\\[4pt]
M_s \sim 10^{17}\,\mathrm{GeV}: \quad
& g_N \sim 10^{-3}, \qquad
g_{b\gamma}^2 \sim 6.7, \qquad
g_N g_{b\gamma}^2 \sim 10^{-3},
\\[4pt]
M_s \sim 10^{16}\,\mathrm{GeV}: \quad
& g_N \sim 10^{-5}, \qquad
g_{b\gamma}^2 \sim 11.6, \qquad
g_N g_{b\gamma}^2 \sim 10^{-3}.
\end{align}
We work in a controlled small-mixing approximation, retaining the leading axion-photon and pure gravitational contributions to the anomalous dimensions, while neglecting mixed terms of order $g_{N}g_{mix}^2$.  

We stress, for the benefit of the reader, that our string-model independent (KR) axion $b(x)$, which enters the analysis of \cite{Mavromatos:2024ozk} and the current article, is massless. 
Such massless KR axions play the r\^ole of totally-antisymmetric torsion degrees of freedom in string theory~\cite{Duncan:1992vz}, to the order in target-space derivative expansion of the effective action we are working on. Thus, the standard axion phenomenology/cosmology~\cite{Marsh:2015xka}
does not apply. 

However, mechanisms for the KR axion to acquire a mass during the post-inflationary epoch, do exist within the StRVM~\cite{bms2,Mavromatos:2021hai}. We remark that the presence of Landau-pole-like singularity in the axion couplings persists in the massive axion case~\cite{Eichhorn:2012uv}, and thus, the conjecture of \cite{Mavromatos:2024ozk} could apply. However, the fixed-point structure of the massive axion-electrodynamics is different in the presence of axion mass.

\section{Conclusions and Outlook}\label{sec:concl}

  In the current article we have extended the FRG study of flat spacetime (massless) axion electrodynamics by including gravitational interactions, but not gravitational anomalies. The model fits within the StRVM framework, a string-inspired framework for cosmology, which yields a running-vacuum model  for the evolution of the Universe. Such models, although consistent with the vast majority of the large-scale cosmological data available at present,  predict observable deviations from $\Lambda$CDM cosmological framework. The deviations  are capable of alleviating the persisting cosmic tensions~\cite{CosmoVerseNetwork:2025alb} in the data measuring the Hubble parameter but also the matter growth at modern eras~\cite{rvm2,Gomez-Valent:2023hov}. 

  In our model for massless axions interacting through topological terms with Abelian gauge fields and Einstein gravity,  we considered truncations in the York decomposition which allowed us to keep within  a small set of couplings. In particular we restricted ourselves to project onto $\eta_{TT}$. In addition, of course, there are issues of ignoring higher curvature and in general spacetime-derivative terms. An effective string-inspired theory of gravity consists formally of an infinity of such terms, and their resummability is of course beyond control, at least at present. Such terms complicate enormously the FRG analysis in the presence of gravity.
  
  Fortunately, in our context of  StRVM cosmology~\cite{bms,bms2,ms1,ms2}~and in the absence of dynamical dilatons, the quadratic terms in Riemann curvature 
  can always be cast in a ghost-free Gauss-Bonnet (GB) form (within the context of perturbative string theory~\cite{Gross:1986mw,Metsaev:1987zx} using the matching of S-matrix amplitudes and the requirements of world-sheet conformal invariance). The GB terms are total derivatives in (3+1)-dimensions, hence for constant dilatons, and up to fourth order in spatial derivatives in the string effective low-energy actions, the only non-trivial higher curvature terms are the CS anomaly terms \eqref{e1}. In the absence of gravitational anomalies, due to their cancellation in the post-inflationary Universe in the StRVM framework, one is left essentially with axion electrodynamics models in the presence of gravity, \eqref{sea4}, as we discussed in Appendix \ref{sec:appstRVM}; chiral U$(1)$ global anomalies remain in this framework. The entire cosmological evolution of the StRVM Universe, from inflation to the current era, can be sufficiently described (from a phenomenological viewpoint) by restricting ourselves to terms of at most fourth-order in derivatives~\cite{ms1,ms2,Dorlis:2024yqw,Gomez-Valent:2023hov}. Hence we need not consider higher order terms in our analysis to formulate our conjecture on the \cPT-symmetric repulsive phase interpretation of the observed acceleration of Universe~\cite{Mavromatos:2024ozk} within the CS theory. 
  
  In the presence of non-constant dilatons, quadratic Riemann curvature terms should be included, which would complicate the FRG analysis. 
   Furthermore the inclusion of a pseudoscalar potential
   has also been ignored since shift symmetry is crucial for restricting ourselves to an Abelian gauge anomaly term coupled to the axion; this term does not contain instantons and the maintenance of the shift symmetry is guaranteed. 
For generic UV complete effective theories of gravity outside the framework of StRVM, 
 more complete truncations would be needed in order to support the conjecture of \cite{Mavromatos:2024ozk}.

The conjecture of a $\cPT$-symmetric, repulsive gravity phase interpretation of the currently observed acceleration of the Universe, if true, is consistent with the swampland conjecture~\cite{swamp1,swamp2,swamp3,swamp4,swamp5} on 
the incompatibility of string theory with the $\Lambda$CDM framework (characterised by  a positive cosmological constant). Moreover, for such a novel interpretation, the recent data~\cite{DESI:2024mwx,DESI:2025zgx,DESI:2025fxa}, pointing to a non-constant relaxing dark energy, 
as well as other cosmological data~\cite{Planck:2018vyg,CosmoVerseNetwork:2025alb}, 
including gravitational-wave astronomy~\cite{Bailes:2021tot}, need to be rethought, and reanalysed. The fact that in our approach, the early (inflation) and late acceleration phases of the Universe are due to entirely different physical mechanisms is intriguing, especially because of the involvement of $\cPT$~symmetric non-Hermitian physics involved in the late cosmological eras. 
This calls for entirely novel interpretations of the available data, 
and perhaps unconventional probes as well.
In this work we do not deal with such important phenomenological issues, which are postponed for future works.

We stress that the massless KR axion discussed in \cite{Mavromatos:2024ozk} and here, plays the role of torsion in spacetime geometry. We also remark that, although the Landau-pole-like singularities in the axion electrodynamics can persist in the \emph{massive-axion} case, and thus the conjecture of \cite{Mavromatos:2024ozk} might be valid, nonetheless the fixed-point structure, and the incorporation of gravity in the pertinent FRG studies, is modified in a non-trivial way. We hope to be able to return to this case in a future work.

There are two different `run with scale' ideas being conflated in the literature. In EFT of gravity one can compute loop corrections and define a scale-dependent $G(\mu)$, but its running is scheme/process dependent and not directly an observable. In the framework of asymptotic safety, $k$ is identified as a coarse-graining scale, which facilitates the study of the flow of the effective average action. The central object is the dimensionless Newton coupling. We comment on a potential contribution of our work to this ongoing debate on the RG running of the
(dimensionful) gravitational constant with a scale. (For a selected partial literature on this see \cite{Anber:2010uj,Ellis:2010rw,Ellis2011-td,Donoghue:2024uay,Eichhorn:2017ylw,Eichhorn:2018whv}).
In the context of asymptotic safety in gravity \cite{Reuter:2019byg,Niedermaier2006-yu}, the running of Newton’s “constant” $G$ is with respect to a
momentum scale that is assumed physical. For asymptotically-safe extensions of the
Standard Model of particle physics,
it is claimed \footnote{See e.g. \cite{Saueressig:2024ojx,Eichhorn:2025sux} and references therein} that on allowing $k$ to probe trans-Planckian momenta (see~\cite{Souma:1999at,Reuter:2001ag,Litim:2003vp,Codello:2008vh,Meibohm:2015twa,Sen:2021ffc}
 for example), the existence of a fixed point  implies predictivity: only finitely many relevant directions correspond to free parameters, while the remaining couplings are fixed by the requirement of reaching the fixed point.  This is also the case in \cite{Eichhorn:2012uv} and in our current work; one can see which couplings of the model are
free parameters, and which can be predicted; one can attribute a physical meaning to the gravitational
contributions to this type of running. In this way, it is argued that the objections of \cite{Anber:2010uj,Donoghue:2024uay} on a potential
physical running of $G$, which essentially pertain to conventional RG-scale-running below the Planck scale,
can be avoided.

The arguments in \cite{Ellis:2010rw,Ellis2011-td,Donoghue:2024uay} against gravitational contributions to gauge couplings, suggested in \cite{Robinson:2005fj}, and elaborated further in  \cite{Toms:2010vy,Toms2011-jd}, relate to the fact that such a running is due to contributions of the effective action, which can be eliminated by appropriate field redefinitions. These leave the physical S-matrix of the theory invariant, 
and hence do not contribute to physical quantities. 
In our approach, the above issues are not relevant
for the main conclusion, namely the conjecture of a phase transition of the Universe to a \cPT-symmetric phase in the infrared. Here we have discussed the contributions of gravity to an FRG study of axion electrodynamics, in which the running of couplings is with respect to a renormalization-group/cut-off scale $k$, within an asymptotic safety scenario. However, for our purposes here and in \cite{Mavromatos:2024ozk}, the issue of a physical running of $G$, which lies at the heart of the aforementioned debate, is not of central importance. The central issue of our approach is the presence of  {\it singularities of the RG flow} of a flat-spacetime dimensionful coupling, the axion coupling. As we have argued above, this singularity structure of flat-space axion electrodynamics remains unaffected by the presence of gravity.

In the current article and in \cite{Mavromatos:2024ozk}, we have identified the infrared energy scale at which the dimensionful axion-coupling RG flow becomes infinite: a phase transition to a \cPT-symmetric phase occurs, at the physical cosmological length scale at which the Universe enters an accelerated expansion. At such points of the non-perturbative FRG flow, the corresponding RG $\beta$-function of the axion coupling diverges. The (renormalised) scattering S-matrix of the model breaks down due to this singularity, even at flat spacetimes. In this case, the Landau pole in the RG flow prompted us in \cite{Mavromatos:2024ozk} to conjecture the continuation of the theory to a new phase with \cPT symmetry and repulsive gravity. This phase has well-defined RG flows for the axion couplings in the infrared regime of physical momenta. The repulsive nature of gravity in the \cPT-symmetric phase is actually independent of the question as to whether a running of $G$ is physical or not  (which lies at the heart of the debate discussed in, {\it e.g.} \cite{Anber:2010uj,Donoghue:2024uay}). This also does not invalidate the arguments of \cite{Ellis:2010rw,Ellis2011-td}, given that the S-matrix in our case is ill defined in the \cPT-broken phase at the singularities of the axion-coupling non-perturbative FRG flow. The singularities are deemed to have  physical significance. Since this work relies on the conjecture of \cite{Ai:2022csx}, it is necessary to investigate further the nature of the analytic continuation across the singular surface encountered in coupling space, the \cPT-metric that replaces standard Hermitian inner product and the gauge/diffeomorphism consistency of the continued theory. These are all nontrivial issues. The robustness of our findings, implied by this naive gravitational calculation, makes this an interesting and necessary program.

\begin{acknowledgments}
NEM thanks the University of Valencia and its Theoretical Physics Department for a visiting Research Professorship
supported by the programme  \emph{Atracci\'on de Talento}
INV25-01-15. 
The work of Leqian Chen is supported by a PhD scholarship from the King's-China Scholarship Council.
The work of NEM and SS is supported in part by the UK Engineering and Physical Sciences Research Council (EPSRC) and  Science and Technology Facilities research Council (STFC) 
under the research grants no.  EP/V002821/1 and ST/X000753/1, respectively. 
NEM also acknowledges participation in the COST Association Actions CA21136 “Addressing observational
tensions in cosmology with systematics and fundamental physics (CosmoVerse)” and CA23130 ”Bridging high and low
energies in search of Quantum Gravity (BridgeQG)”.

\end{acknowledgments}
\vskip .2cm
\appendix
\vskip .1cm

\section{String-inspired Running Vacuum Model (StRVM) of Cosmology: brief review}\label{sec:appstRVM}

We review briefly the main features of the String-inspired Running Vacuum Model of cosmology (StRVM)~\cite{bms,bms2,ms1,ms2}, which \emph{constrains the effective action},  on which the analysis in \cite{Mavromatos:2024ozk} is reliant. For the post-inflationary-era we emphasise the  \emph{cancellation} mechanism of the primordial \emph{gravitational anomalies} that characterise the model. Before arriving at this point, we feel like reviewing briefly some important features of Chern-Simons (CS) gravity, which is the framework in which our approach~\cite{Mavromatos:2024ozk} is developed.

 Our string-inspired model is 
based on the massless gravitational multiplet of superstrings~\cite{str1,str2}, consisting of gravitons, dilatons and antisymmetric (Kalb-Ramond (KR)) fields, the latter being dual in (3+1)-dimensions to the so-called string-model independent axion, or KR axion $b(x)$.\footnote{The duality is expressed via \eqref{duality}, where the three-form $H$ is the field strength of the spin-1 KR field $B_{\mu\nu}=-B_{\nu\mu}$~\cite{str1,str2}, modified by the Green-Schwarz anomaly-cancellation counterterms~\cite{Green:1984sg}, see \eqref{csterms} and subsequent discussion.} 
For constant dilatons, which is the basic assumption of the StRVM,\footnote{In realistic string compactifications (e.g. flux compactifications) the dilaton is often stabilised  by fluxes or nonperturbative effects such  as gaugino condensation. Stabilisation implies that that the dilaton sits at a minimum of an effective potential.} 
and keeping for the moment gauge fields (which, for strings~\cite{str1,str2} are in general non-Abelian (denoted by bold-face field strengths)), 
the pertinent action, that described the dynamics in the early Universe, reads~\cite{Duncan:1992vz}:
\footnote{Our conventions and definitions used throughout this work are: signature of metric $(+, -,-,- )$, Riemann Curvature tensor 
$R^\lambda_{\,\,\,\,\mu \nu \sigma} = \partial_\nu \, \Gamma^\lambda_{\,\,\mu\sigma} + \Gamma^\rho_{\,\, \mu\sigma} \, \Gamma^\lambda_{\,\, \rho\nu} - (\nu \leftrightarrow \sigma)$, Ricci tensor $R_{\mu\nu} = R^\lambda_{\,\,\,\,\mu \lambda \nu}$, and Ricci scalar $R_{\mu\nu}g^{\mu\nu}$. We also work in units $\hbar=c=1$.}
\begin{equation}
	S_{b,\rm grav}=\int d^4x \,\sqrt{-g}\,  \left[\frac{R}{2\kappa^2} +\frac{1}{2}(\partial_\mu b)(\partial^\mu b) -\frac{1}{4}\mathbf F_{\mu\nu}\, \mathbf F^{\mu\nu}     
    + S_{\rm CS} +  \dots \right]\,,
\label{sea1}
\end{equation}
where $g$ is the determinant of the metric $g_{\mu \nu}$, and 
$S_{\rm CS}$
is a four-spacetime-derivative topological Chern-Simons term~\cite{Duncan:1992vz}:\footnote{The interaction terms \eqref{e1} arise in string theory from the Green-Schwarz anomaly-cancellation modification~\cite{Green:1984sg} of the antisymmetric tensor field strength $H_{\mu\nu\rho}$ by appropriate Lorentz (gravitational) $\Omega_{{\rm 3L}} $ and gauge $\Omega_{\rm 3Y}$ Chern-Simons (CS) three-forms (see Appendix \ref{sec:PTanomcanc} for  details) :
\begin{align}\label{csterms}
H (x) &= \mathbf d\,B (x) + \frac{\alpha^\prime}{8\, \kappa} \, \Big(\Omega_{\rm 3L}(x) - \Omega_{\rm 3Y}(x) \Big),  \nonumber \\
\Omega_{\rm 3L} &= {\omega}^a_{\,\,c} (x) \wedge d\,{\omega}^c_{\,\,a} (x) 
+ \frac{2}{3}  {\omega}^a_{\,\,c} (x) \wedge {\omega}^c_{\,\,d} (x) \wedge {\omega}^d_{\,\,a} (x),
\quad \Omega_{\rm 3Y} (x) = \mathbf A (x) \wedge  \mathbf d\,\mathbf A (x) + \frac{2}{3} \, \mathbf A (x) \wedge \mathbf A (x) \wedge \mathbf A (x).
\end{align}
For coordinate independence, we use the language of differential forms~\cite{Eguchi:1980jx},
with $\wedge$ denoting an exterior product among appropriate one forms.
The one-form ${\omega}^a_{\,\, b}(x)$ denotes the torsion-free spin connection  (Latin indices $a,b,c=1,\dots 4$ denote tangent-space indices at a spacetime point $x$), and $  \mathbf A (x)$ are generic non-Abelian gauge-field one-forms that characterize strings. The $\Omega_{\rm 3L} \, (\Omega_{\rm 3Y})$ are  
Lorentz (Yang-Mills) CS three-forms~\cite{Eguchi:1980jx}. The field $b(x)$ arises in a path-integral formalism~\cite{Duncan:1992vz}, as a pseudoscalar Lagrange multiplier, associated with the implementation of the Bianchi-identity constraint. This constraint is derived on taking the exterior derivative on both sides of \eqref{csterms}, which in component form reads:
 \begin{align}\label{modbianchi2}
& \varepsilon_{abc}^{\;\;\;\;\;\;\mu}\, {H}^{abc}_{\;\;\;\;\;\; ;\mu} 
 =  \frac{\alpha^\prime}{32\, \kappa} \, \sqrt{-g}\, \Big(R_{\mu\nu\rho\sigma}\, \widetilde R^{\mu\nu\rho\sigma} -
\mathbf F_{\mu\nu}\, \widetilde{\mathbf F}^{\mu\nu}\Big) \,,
\end{align}
 with the semicolon ($;$) denoting the gravitational covariant derivative with respect to the torsion-free connection. The interaction \eqref{e1}, emerges after path-integrating out the field strength $H_{\mu\nu\rho}$ with respect to the action \eqref{sea1}.} 
\be
\label{e1}
S_{\rm CS} = 
A
\int d^4x\;\sqrt{-g}\;b(x)\,
\Bigl(
R_{\mu\nu\rho\sigma}\,\widetilde R^{\mu\nu\rho\sigma}
\;-\;
\mathbf F_{\mu\nu}\,\widetilde{\mathbf F}^{\mu\nu}
\Bigr)\,, \,
\ee
with 
\begin{align}\label{Adef}
A \equiv \sqrt{\frac{2}{3}}\,\frac{\alpha'}{96\,\kappa}\,.
\end{align}
Here $\kappa = M_{\rm Pl}^{-1}$ (in units $\hbar=c=1 $ ) is the (3+1)-dimensional gravitational coupling (inverse of the reduced Planck mass $M_{\rm Pl} = 2.4 \cdot 10^{18}~{\rm GeV}$),  $\alpha^\prime = M_s^{-2}$ is the \emph{Regge slope} of the string~\cite{str1} and  $M_s$ is the string mass scale. The axion-like particle (ALP) $b(x)$ is independent of the model used for compactification ~\cite{Svrcek:2006yi}.\footnote{ A string-model independent axion is distinct from compactification ALPs, which are abundant in string theory, and depend on details of the string model and its compactification. In the current article we shall restrict our attention to the KR axion $b(x)$ field, and its anomalous coupling \eqref{e1}.}
In \eqref{e1} the symbol $\widetilde{\dots}$ 
denotes the dual of the gauge field strength  and Riemann curvature, respectively:
\begin{equation}
\label{dualriem}
\widetilde{F}_{\mu\nu} = \frac{1}{2} \varepsilon_{\mu\nu\rho\sigma}\, F^{\rho\sigma} \,, \qquad 
\widetilde{R}_{\alpha\beta\gamma\delta}=\frac{1}{2}R_{\alpha\beta}^{\,\,\,\,\,\,\,\,\,\rho\sigma}\varepsilon_{\rho\sigma\gamma\delta}\,,
\end{equation}
with $\varepsilon_{\mu\nu\rho\sigma}$ the covariant Levi-Civita tensor~\cite{Eguchi:1980jx}. The gauge field strength in \eqref{sea1}, \eqref{e1} could be non-Abelian, but in our considerations below, and in \cite{Mavromatos:2024ozk}, we take them to be Abelian, associated with electromagnetic interactions, for simplicity. 
The gravitational theory \eqref{sea1}, \eqref{e1} is a CS gravity~\cite{Jackiw,Alexander:2009tp}.
In effective low-energy limits of string theories, the Lagrangian \eqref{sea1}
is part of an infinite set of (higher mass-dimension) terms, which play a role, dependent on the energy scale under consideration.

Despite their  similarities in form language, the two anomaly terms in \eqref{e1} are different in their contributions to the stress energy tensor $T_{\mu\nu}$. The gauge CS term (last term on the right-hand side of the equality \eqref{e1}) is topological since it is independent of the spacetime metric. Hence its contribution to $T_{\mu\nu}$ vanishes. In contrast, the gravitational CS (gCS) term (first term on the right-hand-side of the equality \eqref{e1}) gives a non-zero contribution to the stress tensor, a Cotton-like symmetric but traceless term $\mathcal C_{\mu\nu}$~\cite{Jackiw}:
\begin{equation}
\label{cottdef}
	C_{\mu\nu}=-\frac{1} {2}\left[(\partial^{\beta} b) \widetilde{R}_{\alpha\mu\beta\nu}+(\partial^{\beta} b) \widetilde{R}_{\alpha\nu\beta\mu}\right]^{;\alpha}~, 
\end{equation}
which yields the following modified Einstein (graviton) equation of motion:
\begin{align}
	\label{grav}
	&G_{\mu\nu}=\kappa^2 T^{(b)}_{\mu\nu} - 2 \kappa^2 A\, C_{\mu\nu} \equiv \kappa^2 T^{\rm improved}_{\,\,\mu\nu}\,.
\end{align}
Here $T^{(b)}_{\mu\nu}$ is the  stress energy-momentum tensor associated with the kinetic term of the massless axion-KR field in the action $T^{(b)}_{\mu\nu}=\nabla_\mu b\nabla_\nu b-\frac{1}{2}g_{\mu\nu}(\nabla b)^2$.
The property of the Cotton-like tensor~\cite{Jackiw}: $C_{\mu\nu}^{\quad ;\mu} = \frac{1}{8} \, (\partial_\nu b)\, R^{\alpha\beta\rho\sigma}\, \widetilde R_{\alpha\beta\rho\sigma}$, implies that the naive covariant conservation of the axion-matter stress tensor fails, due to a non-trivial exchange of energy between the axion and the gravitational anomaly. This does not imply any pathological behaviour of the underlying gravity theory, since diffeomorphism invariance is kept intact, due to the (covariant) conservation of the improved stress-tensor $T^{\rm improved}_{\,\mu\nu}$.

As already mentioned, the conjecture in  \cite{Mavromatos:2024ozk} is motivated by the StRVM framework, in which the dynamics of the primordial, pre-inflationary Universe is described {\it only} by  dilaton, graviton and antisymmetric tensor fields. No other fields (except perhaps instanton configurations of non-Abelian primordial gauge field that appear in the string-theory spectrum~\cite{Dorlis:2024uei}, which, however, are not relevant for our purposes here) are present as external fields in this primordial, pre-inflationary phase. Thus, the pertinent action for this epoch is given by \eqref{sea1} upon setting 
\be
\mathbf F_{\mu\nu}=0 \,. 
\ee
Inflation in this theory, which is of Running Vacuum Model type~\cite{Lima,Perico,rvm1,rvm2}, is due to condensates of the gravitational  CS terms \eqref{e1}, due to chiral primordial Gravitational Waves (GW). The latter imply a vacuum energy density proportiomnal to the fourth power of the Hubble parameter $H^4$ of the primordial string Universe.
The condensate has imaginary parts~\cite{Dorlis:2024uei}, which imply a finite-life-time inflation. To obtain phenomenologically relevant durations of inflation, of order~\cite{Planck:2018vyg} $\mathcal O(50-60)$ e-foldings one needs~\cite{Dorlis:2024uei} $M_s \sim 0.2 \, M_{\rm Pl}$.

Matter and gauge fields are generated by the decay of the RVM unstable vacuum at the exit phase of inflation. Matter includes massless chiral fermions which are assumed such that they cancel the primordial gravitational anomalies. 
The generation of chiral fermionic matter, though, will generate its own gravitational and gauge anomalies, via non-conservation of the chiral current $ J^{5\, \mu} = \sum_i \overline \psi_i \gamma^\mu \, \gamma^5 \, \psi_i $, where $i$ is an index running over the chiral fermionic species $\psi_i$ of the respective string effective theory.

Thus, the effective gravitational theory at the end of the RVM inflationary phase, will be schematically given by~\cite{bms}: 
\begin{align}\label{bplusf}
S_{\rm total} = S_{b,\rm grav} + 
S_{b,F} + \dots 
\end{align}
where $S_{b,\rm grav}$ is given in \eqref{sea1}, and 
the chiral fermionic part of the action $S_{b,F}$ interacting with the $b$ KR axion reads (we suppress the chirality index in the fermions for notational brevity)~\cite{deCesare:2014dga,bms}:
\begin{align}\label{bfaction}
S_{b,F} =  
S_{F,\rm kin} + \int d^4 x\, \sqrt{-g} \, \frac{\kappa}{2\,}\, \sqrt{\frac{3}{2}}\, b \, J^{5\, \mu}_{\,\,\,\,\,\,\,\,\,;\,\mu} + \dots \,,
\end{align}
where $S_{b,\rm kin}$ are the fermion kinetic terms in the presence of torsion-free gravity, $;$ is the torsion-free gravitational covariant derivative, and the $\dots$ denote four-fermion $J_\mu^5J^{5\, \mu}$ terms.\footnote{The interaction term with its specific coupling arises from the fact that to this order in derivatives, the (totally antisymmetric in its covariant indices) field strength $H_{\mu\nu\rho}$ of the antisymmetric tensor field plays the r\^ole of torsion, with a generalised connection~\cite{Metsaev:1987zx,Duncan:1992vz}
\begin{align}\label{torsion}
{\widetilde \Gamma}^\mu_{\,\,\nu\rho} = \Gamma^\mu_{\,\,\nu\rho} + \frac{\kappa}{\sqrt{3}}\, g^{\mu\sigma}\, H_{\sigma\nu\rho} \,,
\end{align}
with $\Gamma^\mu_{\,\,\nu\rho}$ the torsion-free Christoffel connection. The coupling of $b$ to chiral fermions in \eqref{bfaction} is precisely dictated by this r\^ole, after the appropriate dualisation procedure,
{\it i.e.} implementing the Bianchi identity \eqref{modbianchi2} in the presence of fermions, and passing to the dual picture in terms of the KR axion field $b$, during which the $J^5-J^5$ also arise~\cite{deCesare:2014dga}. Such four-fermion terms are characteristic of theories with torsion~\cite{Duncan:1992vz,Mavromatos:2023wkk}. Appropriate integration by parts has also taken place, in order to cast the $b$-fermion interactions in the form \eqref{bfaction}~\cite{deCesare:2014dga}.}

For late eras that we are interested in \cite{Mavromatos:2024ozk} and here, it suffices to assume that only the U(1) Abelian gauge anomaly term coupled to the axion $b$ field remains in the effective axion. Heuristically, which suffices for our purposes, on assuming $N_f$ chiral fermionic degrees of freedom, we would obtain in the low-energy field theory the (one-loop exact) anomaly equation~\cite{Alvarez-Gaume:1983ihn,Morozov:1986hv}:
\begin{align}
J^{5\, \mu}_{\,\,\,\,\,\,\,\,\,;\,\mu} = \frac{N_f}{96\, \pi^2}\, R_{CS} + \frac{e^2}{8\pi^2} F_{\mu\nu} \, {\widetilde F}^{\mu\nu}\,,
\end{align}
where $e$ is the electron charge. 

Consistency of the {\it post-inflationary theory}, which is of interest to us in \cite{Mavromatos:2024ozk} and here, requires~\cite{bms2,bms2,ms1,ms2} that at most chiral global anomalies may remain uncancelled. 
Cancellation of gravitational anomalies in \eqref{bplusf}, 
then, would require
\begin{align}\label{condanomcanc}
N_f = 192\, \pi^2\, \sqrt{\frac{8}{3}}\,\frac{A}{\kappa} = \frac{16}{3}\, \pi^2 \, \Big(\frac{M_{\rm Pl}}{M_s}\Big)^2\,,
\end{align}
where in the last equality we used \eqref{Adef} for the case of string theory.\footnote{It is understood that the choice of appropriate string scales $M_s$ that satisfy \eqref{condanomcanc} are such that $N_f$ is a positive integer.} As mentioned above, for the condensate-induced-inflationary model of \cite{Dorlis:2024yqw,Dorlis:2024uei}, we have $M_s/M_{\rm Pl} \sim 0.2$, which, on account of \eqref{condanomcanc}, would imply an $N_f \sim {\mathcal O}(10^3)$, in agreement with standard string theory models, used also in gravitational leptogenesis scenarios~\cite{Alexander:2004us}.\footnote{We remark, for completion, that StRVM also involves post-inflationary  gravitational leptogenesis, via unconventional scenarios, however, entailing spontaneous Lorentz and CPT violation~\cite{deCesare:2014dga,bms}, as a consequence of the gCS condensates during the RVM inflationary era. Thus, StRVM stands a chance of providing a phenomenologically realistic cosmology.}

The late eras cosmology of StRVM, then, would encompass a standard gCS-anomaly-free axion electrodynamics for the KR field $b(x)$ in curved space time:
\begin{align}
	S_{b,A_\nu,\rm grav} =\int d^4x \,\sqrt{-g}\,  \left[\frac{R}{2\kappa^2} + \frac{1}{2}(\partial_\mu b)(\partial^\mu b) - \frac{1}{4} F_{\mu\nu}\, F^{\mu\nu} + A\, b F_{\mu\nu}\, {\widetilde F}^{\mu\nu} \right]\,,
\label{sea4}
\end{align}
where $A_\mu$ the photon field, and the coefficient $A$ is given in \eqref{Adef}. 

The Euclidean version of the effective theory \eqref{sea4} is the basic theory which was used in \cite{Mavromatos:2024ozk} and here, to formulate the conjecture that the observed late-era acceleration of our Universe corresponds to a phase transition to a \cPT symmetric, but repulsive phase of gravity. Under the action of a FRG, the coupling $A$ develops singularities in its Renormalizastion-Group (RG) flow in the infrared, 
in the presence of gravity, handling of 
which lead to the aforementioned phase transition, with the \cPT-symmetric theory being characterised by finite FRG flows in the infrared, but with matter in this theory feeling a repulsive gravity. This leads to the \cPT-symmetric-universe acceleration at late eras (large cosmological (infrared) length scales), which in \cite{Mavromatos:2024ozk} has been conjectured to correspond to the observed accelerated expansion phase of our Universe, thereby alleviating the need for the assumption of a dominant dark energy component.

\section{On \cPT ~Symmnetry : Basic features}\label{sec:PT}

Global discrete $\cPT$ symmetry is mostly considered in the context of flat spacetime, but classically it naturally translates, to curved spacetime \footnote{We will be considering gravity linearised around a flat background. Minkowski (Lorentzian) manifold does allow charts and an atlas, which is a notion related to topology and smoothness  }. Parity $\mathcal P$ is a discrete diffeomorphism of the spacetime manifold $M$ with a composition $\mathcal P \circ \mathcal P=id_{M}$, where $id_{M}$ is the identity mapping. Morover $\mathcal P$ is an isometry of the metric. As usual we  get a familiar representation of $\mathcal P$ by choosing charts and an atlas for $M$. In a chart we choose $x: M \rightarrow \mathbb{R}^{4}$ and $\mathcal P: x^{\mu} \rightarrow \bar{x}^{\mu}=\mathcal P^{\mu}(x)$; in a particular case we may have $\mathcal P^{\mu}{(x)}=\left(x^{0},-x^{1},-x^{2},-x^{3}\right)$. The operator $\mathcal T$ is handled in similar way (provided the manifold is time-orientable). In  \emph{quantum theory} $\mathcal T$ is an antilinear operator on the space of fields. In flat spacetime with Lorentzian signature $(p, q)=(3,1)$, the real orthonormal-frame group $O(3,1)$ has $4$ disconnected pieces (its cosets) 
  labelled by orientation and time-orientation. In Euclidean signature $(p, q)=(4, 0)$ the frame group is $O(4)$ and has only proper or improper rotations. Although in general relativity the vierbein bundle is a principal $O(3,1)$ bundle \cite{Nakahara2018-on}, in any small enough patch of the atlas, the frame bundle can be trivialised.  Every local $O(3,1)$ transformation is written as $\Lambda(x)=P^{\epsilon_{P}(x)} \mathcal T^{\epsilon_{T}(x)} L(x)$ where $L(x)$ is a \emph{proper orthochronous} transformation;  $P^\epsilon_{P}$ and $T^\epsilon_{T}$ change the space and time orientation for $L(x)$ depending on $\epsilon_{P},\epsilon_{T} \in \{0,1\}.$ 

A new form of  quantum mechanics emerges where the role of Hermiticity is replaced by \cPT ~symmetry \cite{PTqm1}. Non-Hermitian Hamiltonians $H$ satisfying $[H,\cPT]=0$ can, remarkably, have entirely real spectra \cite{Bender_2007,Dorey2001-dj} for a large family of Hamiltonians; the phase of $H$ is determined by the domain of definition of $H$ on a Hilbert space or  the class of allowed paths in a path integral. Just as in the evaluation of the asymptotic behaviour of real integrals, where the method of steepest descents is used involving the complex plane, the path integral considers complex paths using a generalisation (of the method of steepest descents)  known as Lefschetz thimbles \cite{Witten:2010cx}. This has led to much interest in constructing unitary field theories based on $\cPT$ symmetry (see, for example,\cite{Bender_2016,Alexandre:2020gah}) in Euclidean path integrals. The equivalence of canonical and path integral formulations of \cPT-symmetric theories is not settled even though there are many indications that calculations of Green functions can be done using either formulation \cite{Jones:2009br}. In particular  path integral formulations are often simpler to implement. An explicit determination of new inner products associated with \cPT~symmetry is not required  for  calculations in some \cPT-symmetric scalar field theories \cite{Croney:2023gwy}. \cbl Our considerations are for physics beyond the standard model of particle physics (BSM) and \cPT~ symmetry  is  an example of BSM. Such theories are commonly implemented using path integrals in four spacetime dimensions. We assume, for our purposes in this work, that the path integral approach is valid for \cPT~ symmetry. Given this starting point, it becomes possible to investigate the role of renormalisation in \cPT~ quantum field theory and make connections with cosmology, and this is  attempted in \cite{Mavromatos:2024ozk}, 
and here, within the context of the StRVM, reviewed in Appendix \ref{sec:appstRVM}.

 \cPT ~symmetry can arise due to renormalisation in quantum systems \cite{Bender:2021fxa}. This was first noticed in a nonrelativistic toy field theory known as the Lee model~\cite{Lee}. This early model was quantum mechanical and invented  in order to study the phenomenon of renormalisation rigorously (which was facilitated by removing crossing-symmetry). In this model, starting from a Hermitian Hamiltonian, renormalisation forced the development of an imaginary coupling. Before the advent of $\cPT$ symmetry, this model was abandoned as being unphysical. In the important work of Ref.~\cite{Bender:2004sv} it was shown that the introduction of a $\cPT$ inner product led to a reinterpretation of the Lee model in a new $\cPT$-symmetric phase. We shall now give a discussion of the emergence of \cPT~ symmetry in  some simple standard field theoretic contexts \cite{Mavromatos:2022heh}. Such studies are precursors of our work on a gravitational Chern-Simons model. \cite{Mavromatos:2024ozk,Croney:2023gwy}.

\section{$\cPT$~symmetry in Chern-Simons theories}\label{sec:PTCS}

By choosing 4-dimensional diagonal matrices for $\mathcal P$ and $\mathcal T$ (and noting the pseudovector nature of magnetic fields) we find for electrodynamics that under $\cPT$, $E_{i} \rightarrow E_{i}, \quad B_{i} \rightarrow B_{i}$. Hence, under $\cPT$, 
  \be
  F \tilde{F}=\vec{E} \cdot \vec{B} \xrightarrow{\cPT} \vec{E} \cdot \vec{B}\,,
  \ee
  where $F_{\mu\nu}= \partial_\mu A_\nu - \partial_\nu A_\mu$, 
  with $A_\mu$ the U(1) gauge field, and 
  the Hodge dual $\tilde F$ is defined by
\be
\widetilde F^{\mu\nu}
=\tfrac12\,\epsilon^{\mu\nu\rho\sigma}F_{\rho\sigma}.
\ee 
So the interaction term
  \be
\alpha \,b F \tilde{F} \xrightarrow{\cPT} - \alpha \, b F \tilde{F},
  \ee
  for pseudoscalar $b$ and real coupling \cy$\alpha $ \cbl, is Hermitian but not $\cPT$-symmetric. This should be contrasted with the case of pure imaginary coupling $i\alpha$ 
  \be
  i \,\alpha \, b \, F \tilde{F} \xrightarrow{\cPT} i\, \alpha\, b\, F \tilde{F},
  \ee
  which is non-Hermitian but \cPT- symmetric. $b$ is taken to be \emph{even} under $\mathcal T$, which is required when it is the dual of the Kalb-Ramond field strength. We have two possible actions:
  \begin{align}
S_E[\alpha\in\mathbb R]
&=
\int d^4x\,
\biggl[\;
\frac14\,F_{\mu\nu}F_{\mu\nu}
\;+\;i\frac{\alpha}{4}\,b\,F_{\mu\nu}\widetilde F_{\mu\nu}
\biggr],
\\
S_E[\alpha \to i\,\alpha',\;\alpha'\in\mathbb R]
&=
\int d^4x\,
\biggl[\;
\frac14\,F_{\mu\nu}F_{\mu\nu}
\;-\frac{\alpha'}{4}\,b\,F_{\mu\nu}\widetilde F_{\mu\nu}
\biggr].
\end{align}

  In the work of \cite{Mavromatos:2024ozk} it was suggested that singularities in the renormalisation group flow \cite{Croney:2023gwy,Chen:2024ynx} and the requirements of a finite partition function led to a phase change from Hermitian to \cPT~phase. There is an analogue with conventional phase transitions, where for finite volume the partition function $Z$ is analytic in couplings. In the thermodynamic/continuum limit it splits into branches (phases) selected by different saddles/contours (Lefschetz thimbles). In \cPT~symmetric models the same phenomenon appears as Stokes transitions; as couplings are varied the set of thimbles that makes the nonHermitian but \cPT-invariant path integral jumps. Heuristically the renormalised theory defines $\beta \left( g \right) =\mu \frac{\partial g}{\partial \mu}$  from renormalisation functions built out of Green's functions. Here we recall the argument of \cite{Jones:2009br,Croney:2023gwy}, which states that Green's functions can be calculated using path integrals without including any additional measure factor. On approaching a Stokes line certain counter terms can diverge and diagrammatic resummations switch dominant saddles (or thimbles). $Z$ exists on \emph{both} sides but the dominant thimbles (and thus the effective free-energy branch) differ. The Stokes set (singularities)  is where these thimbles reconnect or rearrange. In finite volume this is perfectly smooth; in the limit of infinite volume limit the rearrangement is nonanalytic. 
  These arguments are framed in Euclidean space. \cbl 
  
  Moreover similar  considerations apply if the Abelian $F\tilde{F}$ is replaced by the nonAbelian or gravitational analogues. These theories are  conventionally nonrenormalisable and the set of possible interactions in  theory space, which are connected by renormalisation, is arbitrarily large, unless there are energy-scale, symmetry or topological reasons for restricting the set. 
  
  Next we will show how the $\cPT$-symmetric version arises naturally in (massless) axion physics. 
  
We first define the theory in Minkowski signature  $(+,-,-,-)$. For a massless pseudoscalar $b$,
\begin{equation}
\mathcal L_M \;=\; -\frac{1}{4}F_{\mu\nu}F^{\mu\nu}
\;+\;\frac{g_{mix}}{4}\,b\,F_{\mu\nu}\tilde F^{\mu\nu},\qquad
\tilde F^{\mu\nu}=\tfrac12\,\varepsilon^{\mu\nu\rho\sigma}F_{\rho\sigma}.
\end{equation}
In the StRVM model~\cite{bms,ms1,ms2}, which the conjecture of \cite{Mavromatos:2024ozk} is based upon, one has ({\it cf.} \eqref{e1}):
\be\label{gmixA}
g_{\rm mix} = \sqrt{\frac{2}{3}}\frac{1}{24} \frac{M_{\rm Pl}}{M_s^2}\,.
\ee
With standard $\cP$ and $\cT$ assignments (and $b$ a pseudoscalar),  we recall that 
$\cPT:\, F_{\mu\nu}\tilde F^{\mu\nu}\mapsto -\,F_{\mu\nu}\tilde F^{\mu\nu}$.
Hence, for real $g_{mix}$ the interaction is Hermitian but not \cPT-invariant; if
$g_{mix}\to i{g'}_{mix}$ with ${g'}_{mix}\in\mathbb R$, the interaction becomes non-Hermitian
but \cPT-invariant.

For the FRG we work in Euclidean signature. Under Wick rotation $t\to -i x_4$,
\begin{equation}
S_M=\int d^4x\,\mathcal L_M
\;\longrightarrow\;
S_E=\int d^4x_E\,\mathcal L_E,\qquad e^{-S_E}=\int\mathcal DA_\mu \, \mathcal D b\,e^{iS_M}\!,
\end{equation}
and the Lagrangian reads
\begin{equation}
\mathcal L_E \;=\; +\frac{1}{4}F_{\mu\nu}F_{\mu\nu}
\;+\;\frac{i\,{g}_{mix}}{4}\,b\,F_{\mu\nu}\tilde F_{\mu\nu}\,,
\qquad
\tilde F_{\mu\nu}=\tfrac12\,\varepsilon_{\mu\nu\rho\sigma}F_{\rho\sigma}\,.
\end{equation}
Thus a real Minkowski coupling $\alpha$ becomes an imaginary coefficient $i\alpha$ in $S_E$.
Equivalently, the \cPT~ choice ${g}_{mix}=i{g}_{mix}'$ (Lorentzian) appears with a real coefficient
$+{g}_{mix}'$ in $S_E$.

\section{Motivation for Chern-Simons model  from anomaly cancellation and the role of shift symmetry}\label{sec:PTanomcanc}

The action in \eqref{e1} arises in superstring theory~\cite{str1,str2,pol1,pol2}, after \emph{compactification} to four space-time dimensions; the bosonic ground state of the closed-string sector consists of {\it massless} fields in the so-called {\it gravitational multiplet}, which contains $\phi (x)$ a spin-$0$ (scalar) dilaton , $g_{\mu\nu}(x)$
a spin-2 traceless symmetric tensor field,  identified as the  (3+1)-dimensional graviton, and $B_{\mu\nu}(x)(=-B_{\nu\mu}(x))$ a spin-1 antisymmetric tensor gauge field known as the Kalb-Ramond (KR) field, related to a pseudoscalar axion $b(x)$. We set the four-dimensional dilaton field to be a constant, $\phi(x)=\phi_0$, which fixes the string coupling 
\be\label{stringcoupl}
g_s=\exp(\phi) = \exp(\phi_0)~. 
\ee
Since there are  consistent solutions of  four-dimensional string theory with such a configuration, we shall adopt this treatment of the dilaton. Associated with the KR $B$-field there is a U(1) gauge symmetry of the closed-string (3+1)-dimensional target-space-time effective-field-theory action: 
\be\label{Bgauge}
B_{\mu\nu}(x) \, \rightarrow \, B_{\mu\nu}(x) + \partial_{[\mu}\theta_{\nu]}(x), \quad \mu,\nu =0, \dots 3, 
\quad \theta_\mu (x) \in \mathbb R\,,, 
\ee
where the brackets $[ \dots ]$ denote antisymmetrisation of the respective indices, and a suitable 3-form field strength $\mathcal H$ of the $B$-field \cite{deCesare:2014dga}.

We are also influenced by M theory {\it phenomenology} and anomaly cancellation (in higher dimensional spacetime), for inclusion of Kalb-Ramond fields in orbifold compactifications of M theory. Such compactifications as an essential ingredient in models beyond the standard model (SM) of particle physics . SM has \emph{chiral} fermions, $\psi$ say,  with action in curved spacetime given by 
$S[\psi, e]=\int d^{4{}} x \, e \, \bar{\psi} \,i\, \gamma^{\mu} D_{\mu} \psi$ where $D_{\mu}$ is the spinor covariant derivative and (in terms of vierbeins) $e=\operatorname{det} e^{a}_\mu$. Even though the classical action is invariant under 
local Lorentz transformations and diffeomorphisms, the quantum theory may not be. The reason is that the path integral measure for a chiral fermion is not invariant under such transformations. This leads to an anomalous variation of the effective action $\Gamma$. 
In four--dimensional \emph{chiral} gauge theories triangle diagrams can generate gauge and mixed gauge--gravitational anomalies.  If left uncancelled, these spoil the consistency of the quantum theory.  A compact way to remove such anomalies is to introduce an antisymmetric 2--form field $B_{\mu\nu}$ (the \emph{Kalb--Ramond} field) and couple it through a Green--Schwarz (GS) counterterm.  Below we briefly sketch summarise how this mechanism might work in a stand‐alone effective‐field‐theory setting.

For simplicity we first consider a theory with Abelian and non-Abelian gauge interactions. In the context of anomaly cancellation involving a chiral fermion coupled to a non-Abelian gauge group ${\mathcal G}$, we shall use $T^a$ to refer to the generators of the group $\mathcal G$ in some representation $\mathscr{R}$ , under which the  chiral fermions transform. 

In the SM the gauge group is $SU(3)_{c}\times SU(2)_{L}\times U\left( 1 \right)_{Y}$For a chiral $U(1)_Y$ with fermions of charges $q_i$ the one--loop variation of the effective action reads
\begin{equation}
\delta_{\alpha}\Gamma_{\text{1--loop}} = \frac{1}{24\pi^{2}}\Bigl(\sum_i q_i^{3}\Bigr)\int \alpha\,F \wedge F 
+ \frac{1}{24\pi^{2}}\Bigl(\sum_i q_i\,\mathrm{Tr}T^{2}\Bigr)\int \alpha\,\mathrm{Tr}(G \wedge G),
\label{eq:anomaly}
\end{equation}
where $F$ is the $U(1)_Y$ field strength, $\alpha(x)$ is the infinitesimal parameter of the $U(1)_A$ gauge transformation and $G$ is the field strength for the appropriate non-Abelian group $\mathcal G$.  The two coefficients measure, respectively, the cubic $U(1)_A^3$ and mixed $U(1)_A$--$\mathcal G^2$ anomalies. In SM the assignments $q_i$ are such that the coefficients of the anomalous terms vanish, necessary for the consistency of the theory. However there are additional $U(1)$ axial symmetries\footnote{Heterotic $E_{8} \times E_{8}$ or type-II orientifold models \cite{Ibanez2012-iq} generically yield $4-8$ Abelian generators after compactification. Exactly one linear combination is anomalous. }  in compactifications from higher dimensional theories (to be denoted by $U(1)'$), and these need to be cancelled  in a different way due to Green and Schwarz \cite{Green:1984sg} since the necessary constraints . This new form of cancellation (in higher dimensions) is mathematically complicated and involves the descent equations\cite{Bertlmann2000-na}, which are a mathematical tool used to analyse anomalies in gauge or gravitational theories, specially in the context of cohomological approaches. These equations arise from the observation that gauge anomalies in $d$ dimensions can be related to characteristic classes (like the Chern character) in $d+2$ dimensions. Schematically we start with a closed, gauge invariant form $Tr F^{n+1}$ in $2n+2$ dimensions. Using the descent procedure we write $Tr F^{n+1}=d\Omega^{(0)}_{2n+1}$  and  under a gauge variation 
\be
\delta \Omega_{2n+1}^{\left( 0 \right)} =d\Omega_{2n}^{\left( 1 \right)}
\ee
where $\Omega_{2n}^{\left( 1 \right)}$ is the anomaly polynomial, a local anomalous term. 

Suppose we have a chiral fermion in $4D$, coupled to a a gauge field $A$. Its anomaly polynomial\cite{Alvarez-Gaume:1983ihn,Alvarez-Gaume:2022aak} is a 6-from in 6D:
\be
I_{6}=Tr\left( F^{3} \right).
\ee
This is a closed and gauge-invariant 6-form on a fictitious $6$D space, often called the anomaly polynomial. We now use the descent procedure \cite{Zumino:1983rz} to write $I_{6}=d\Omega_{5}^{\left( 0 \right)}$ and $\delta_{\alpha} \Omega_{5}^{\left( 0 \right)} =d\Omega_{4}^{\left( 1 \right)} \left( \alpha \right)$. For $U(1)$ we have $\Omega_{5}^{\left( 0 \right)} =A\wedge F\wedge F$. Under a gauge variation $\delta A=d \alpha$ and so $\delta \Omega_{5}^{\left( 0 \right)} =d\left( \alpha F\wedge F \right) =d\Omega_{4}^{\left( 1 \right)}$. To cancel the anomaly or reproduce this variation, one introduces a scalar field $b(x)$, which plays the r\^ole of an axion or Green-Schwarz field in 4D. We construct the $4D$ dynamical CS term  
\be
S_{CS}=\int_{M^{4}} b\left( x \right) F\wedge F\ee
or for gravity (in form language, for economy in notation): $S_{CS}^{grav}=\int_{M^{4}} b\left( x \right) Tr\left( R\wedge R \right)$.
 It is thus  plausible that a symmetry based on a shift of  the axion field $b(x)$ can be used to cancel the anomalous terms in \eqref{eq:anomaly} through adding to the action a term 
\be
S_{GS}\sim c_{L}\int b\left( x \right) Tr\left( G\wedge G \right).\ee
Without further justification, we considere for our purposes in this work the r\^ole of shift symmetry in nonrenormalisation of CS couplings in the main text.

\section{\cPT~ symmetry and analytic continuation in String Theory}\label{sec:analcontstring}

In this Appendix, we will argue how the $\cPT$-symmetric version arises naturally in axion physics within the framework of string-in spired effective actions. 
As a result of analytic continuation in the context of string effective theories,  an argument has been made in \cite{mavromicro}, motivating the potential existence of a 
non-Hermitian axion-gravitational-CS-anomaly interaction.  In Euclidean space there is a known ambiguity~\cite{strom} in analytically continuing the CS effective action  back to Minkowski space-time; this stems from properties of the Levi-Civita tensor. 
 In string theory~\cite{str1,str2}, after compactification to (3+1)-dimensional spacetimes, the KR axion is dual to the field strength $H_{\mu\nu\rho}$ of the antisymmetric tensor field $B_{\mu\nu}$. We have fixed the dilaton to be a constant, and so \cite{Mavromatos:2024ozk})~\cite{Duncan:1992vz}: 
\begin{align}\label{duality}
\partial_\mu b = \frac{1}{\sqrt{2}}\, \varepsilon_{\mu\nu\rho\sigma} H^{\nu\rho\sigma} 
\end{align}
To lowest order in derivatives, the dynamics of $B_{\mu\nu}$
(which is a U(1) gauge field in the closed string sector~\cite{str1}, with a gauge transformation \eqref{Bgauge}) is described by the 
kinetic term:
\begin{align}
 S_{\rm H-field} = - \int d^4x \, \sqrt{-g} \, \frac{1}{6} \, H_{\mu\nu\rho}\, H^{\mu\nu\rho}\,.  
\end{align}
Because of \eqref{duality},  the Euclidean path integration over the $H$-field-strength  results in 
an \emph{Euclidean} (E) \cbl string effective action of the form \cite{mavromicro}:
\be\label{square}
S_{\rm eff \, B}^{\rm (E)} \ni \int d^4 x \, \sqrt{g^{\rm (E)}} \frac{1}{12} \varepsilon_{\mu\nu\rho\lambda}^{\rm (E)} \, 
 \varepsilon^{\mu\nu\rho\sigma\, \rm (E)} \, \partial^\lambda b \, \partial_\sigma b,
\ee
with a $b(x)$ term, quadratic in space-time derivatives, and dependent on the covariant Levi-Civita tensor density: 
\begin{equation}\label{leviC}
\varepsilon_{\mu\nu\rho\sigma} = \sqrt{-g}\,  \epsilon_{\mu\nu\rho\sigma}, \quad \varepsilon^{\mu\nu\rho\sigma} =\frac{{\rm sgn}(g)}{\sqrt{-g}}\,  \epsilon^{\mu\nu\rho\sigma},
\end{equation}
 The notation (E) in (\ref{square}) indicates that this quantity is evaluated in a metric of Euclidean-signature. 
 If we {\it first} use the following property of the Levi-Civita tensor density in four dimensional space-time with Euclidean metric:
\be\label{propLC}
\varepsilon_{\mu\nu\rho\lambda}^{\rm (E)} \, 
 \varepsilon^{\mu\nu\rho\sigma\, \rm (E)} = + 6 \, \delta_\lambda^{\, \sigma}~, 
 \ee
where $\delta_\lambda^{\, \sigma}$ denotes the Kronecker delta, and {\it then} analytically continue to Minkowski space-time, we obtain a \emph{real} effective action for the dynamical field $b(x)$ (which plays the r\^ole of the KR gravitational axion). Alternatively, if one {\it first} analytically continues \eqref{square} to a Minkowski-signature space-time, and {\it then} 
uses the Minkowski version of \eqref{propLC} 
\be\label{propLCM}
\varepsilon_{\mu\nu\rho\lambda} \, 
 \varepsilon^{\mu\nu\rho\sigma} = - 6 \, \delta_\lambda^{\, \sigma}~. 
 \ee
The minus sign on the right-hand side is due to the dependence of the contravariant Levi-Civita tensor density  \eqref{leviC}
on the (Minkowski) signature of the metric tensor. 
 The resulting effective action is nonunitary, since the $b$ field has the wrong-sign kinetic term  with respect to the Einstein-Hilbert term in the effective action. In \cite{mavromicro} a field redefinition 
\begin{align}
\label{bib}
b(x) \, \to \,  i\, \widetilde b(x) \,,
\end{align}
was suggested as a way of correcting the wrong-sign kinetic term of the KR axion. However then any gauge/gravitational anomaly terms $S_{CS}$ is non-Hermitian but \cPT -symmetric when written in terms of $\tilde b$. 

Nonetheless, as already stated in the introduction of the current article, this by itself is not sufficient to guarantee continuation of the gravitational coupling \eqref{analcontgrav}. One needs the specific relation of the axion coupling to $\kappa$, 
\eqref{faA}, 
provided in the case of the string-model independent axion~\cite{Duncan:1992vz}. 
It is for this model, which also lies at the heart of the StRVM, for which the conjecture of \cite{Mavromatos:2024ozk} is valid.  The relation (\ref{analcontgrav}) gives context to the conjecture in \cite{Mavromatos:2024ozk} on the emergence of a non-Hermitian \cPT-symmetric phase of $S_{CS}$ 
since we see that the theory in the $\cPT$ phase is a possible outcome of our derivation.

\section{The role of singular flows in simple models}\label{sec:singflows}

  We give simple examples of singular perturbative renormalisation group flows in field theory and their potential connection with \cPT-symmetric phase transitions.
 \vskip .1cm
 Consider a single real scalar field in four dimensions:
\be\label{scalag}
\mathcal L = \tfrac{1}{2}(\partial\phi)^2 
+ \tfrac{1}{2} m^2 \phi^2 
+ \tfrac{\lambda}{4!}\,\phi^4, 
\qquad \lambda > 0.
\ee

For weak coupling in four dimensions the beta function (at oneloop \cite{Croney:2023gwy}) is positive,
\begin{equation}
  \beta(\lambda) \equiv \frac{d\lambda}{d\ln k} = b\,\lambda^2 + \mathcal{O}(\lambda^3),
  \qquad b>0 \quad \big(b=\tfrac{3}{16\pi^2}\ \text{for } \phi^4\big).
\end{equation}
Integrating between a reference scale $\mu$ and $k$ gives
\begin{equation}
  \frac{1}{\lambda(k)} = \frac{1}{\lambda(\mu)} - b \ln\!\frac{k}{\mu}.
  \label{eq:running}
\end{equation}
The coupling blows up at the Landau scale
\begin{equation}
  k_{\mathrm L} = \mu\,\exp\!\left(\frac{1}{b\,\lambda(\mu)}\right).
\end{equation}
Let the \emph{bare} coupling be $\lambda_0 \equiv \lambda(\Lambda)$ at the cutoff $\Lambda$,
and the \emph{renormalized} coupling be $\lambda_R \equiv \lambda(\mu)$ at fixed $\mu$.
Evaluating \eqref{eq:running} at $k=\mu$ and $k=\Lambda$ yields the matching relation
\begin{equation}
  \frac{1}{\lambda_R} = \frac{1}{\lambda_0} + b \ln\!\frac{\Lambda}{\mu}.
  \label{eq:matching}
\end{equation}
Solving for the bare coupling as a function of $\Lambda$ at fixed $\lambda_R>0$,
\begin{equation}
  {\quad
  \lambda_0(\Lambda) = \frac{1}{\displaystyle \frac{1}{\lambda_R} - b \ln(\Lambda/\mu)} \quad}
  \label{eq:lambda0}
\end{equation}
Define the denominator
\begin{equation}
  D(\Lambda) \equiv \frac{1}{\lambda_R} - b \ln\!\frac{\Lambda}{\mu}.
\end{equation}
Then:
\begin{itemize}
  \item For $\Lambda$ close to $\mu$, $D(\Lambda)>0$ and $\lambda_0>0$.
  \item $D(\Lambda)$ crosses zero at $\Lambda = k_{\mathrm L}$:
    \begin{equation}
      D(k_{\mathrm L})=0 \quad\Longleftrightarrow\quad
      \ln\!\frac{\Lambda}{\mu}=\frac{1}{b\,\lambda_R}.
    \end{equation}
    At that point $\lambda_0 \to \pm\infty$.
  \item For $\Lambda>k_{\mathrm L}$, $D(\Lambda)<0$ and hence
    \begin{equation}
      \lambda_0(\Lambda) < 0.
    \end{equation}
    As $\Lambda\to\infty$,
    \begin{equation}
      \lambda_0(\Lambda) \;\simeq\; -\frac{1}{\,b\,\ln(\Lambda/\mu)\,}\ \xrightarrow[\Lambda\to\infty]{}\ 0^{-}.
    \end{equation}
\end{itemize}
Thus, keeping a fixed positive $\lambda_R$ while sending the cutoff beyond the Landau scale
forces the bare coupling $\lambda_0$ to become negative (it does not pass smoothly through zero;
it diverges at $\Lambda=\Lambda_{\mathrm L}$ (the Landau pole\footnote{In the non-perturbative renormalisation group to be considered later the betafunction will be a ratio of two polynomials.}) and reappears negative for $\Lambda>\Lambda_{\mathrm L}$ where $\Lambda_L = \mu_0 \exp\!\left(\frac{16\pi^2}{3 \lambda(\mu_0)}\right)$ ).\cbl

This indicates that ordinary Hermitian $\lambda\phi^4$ theory  may not be UV complete.
If instead $\lambda < 0$, then the same equation shows
\be
\lambda(\mu) \;\to\; 0^- \quad \text{as } \mu\to\infty,
\ee
i.e.\ the theory is \emph{asymptotically free}. 
For $\lambda<0$ the classical potential
\be
V(\phi) = \tfrac{1}{2} m^2 \phi^2 + \tfrac{\lambda}{4!}\phi^4
\ee
is unbounded below, implying that Hermitian quantum theory will have complex eigenvalues.~\cbl
In  \cPT-symmetric quantisation \cite{Dorey2001-dj} using path integrals, one replaces the real $\phi$-integration contour 
with \emph{Stokes wedges} in the complex $\phi$ plane \cite{Bender_2007,Croney:2023gwy}. For the ``wrong-sign'' Hamiltonian
\be
\mathcal H = \tfrac{1}{2}\pi^2 + \tfrac{1}{2}(\nabla\phi)^2 
+ \tfrac{1}{2} m^2 \phi^2 - g \phi^4 \qquad (g>0),
\ee
the Euclidean path integral converges on these rotated contours, and the spectrum 
remains real. In this framework the negative-coupling quartic theory is 
consistent and asymptotically free.

Thus:
\begin{itemize}
\item In the IR ($\mu < \Lambda_L$), the Hermitian $\lambda>0$ theory is valid.
\item Near $\Lambda_L$, the Hermitian description breaks down: the RG trajectory hits the pole.
\item Analytically continuing around the pole lands one on the PT-symmetric 
$\lambda<0$ branch, which is UV safe.
\end{itemize}

Hence the Landau pole acts as a branch point between two analytic sheets: the 
Hermitian $\lambda>0$ theory and its \cPT-symmetric completion with $\lambda<0$.
A similar discussion carries over to gauge theories such as quantum electrodynamics.
Let $\alpha\equiv e^2/(4\pi)$. For $N_f$ Dirac fermions of unit charge, the one-loop beta function is
\begin{equation}
\beta_{\text{QED}}(\alpha)\equiv \mu\frac{d\alpha}{d\mu}
=\frac{2N_f}{3\pi}\,\alpha^2 + O(\alpha^3),
\qquad
\Rightarrow\quad 
\frac{1}{\alpha(\mu)}=\frac{1}{\alpha(\mu_0)}-\frac{2N_f}{3\pi}\ln\!\frac{\mu}{\mu_0}.
\end{equation}
Thus $\alpha(\mu)$ grows with $\mu$ and a UV Landau pole appears upon naive extrapolation.
The \cPT-symmetric version of QED is defined by
\begin{equation}
\mathcal L_{\rm PT}=-\frac14F_{\mu\nu}F^{\mu\nu}
+\bar\psi\,(i\slashed\partial-m)\psi
+ e_{\rm PT}\,\bar\psi\gamma^\mu\psi\,A_\mu,
\qquad e_{\rm PT}\equiv i e,
\end{equation}
with the key discrete assignment that the gauge field transforms as a \emph{pseudovector} under parity:
\begin{equation}
\mathcal{P}:\quad A^0(t,\mathbf x)\to -A^0(t,-\mathbf x),\qquad 
\mathbf A(t,\mathbf x)\to +\,\mathbf A(t,-\mathbf x),
\end{equation}
so that the full Hamiltonian is invariant under the \emph{combined} anti-linear $\cPT$. With $e_{\rm PT}=i e$,
\begin{equation}
e_{\rm PT}^{\,2}=(i e)^2=-\,e^2,
\end{equation}
which will flip signs in loop coefficients built from $e^2$.
The (vacuum-polarization) diagram that determines the one-loop running depends on $e^2$ in ordinary QED. In \cPT-QED,
\begin{equation}
\beta_{\text{PT-QED}}(\alpha_{\rm PT})
=\mu\frac{d\alpha_{\rm PT}}{d\mu}
= -\,\frac{2N_f}{3\pi}\,\alpha_{\rm PT}^{\,2} + O(\alpha_{\rm PT}^{\,3}),
\qquad \alpha_{\rm PT}\equiv \frac{e_{\rm PT}^{\,2}}{4\pi} = -\,\alpha,
\end{equation}
so that
\begin{equation}
\frac{1}{\alpha_{\rm PT}(\mu)}=\frac{1}{\alpha_{\rm PT}(\mu_0)}+\frac{2N_f}{3\pi}\ln\!\frac{\mu}{\mu_0},
\end{equation}
i.e. \cPT-QED is \emph{asymptotically free} at one loop (the screening/antiscreening role is reversed).
\vskip .2cm
 
\section{Effective Average action}
\label{EffAction}

In FRG the standard path integral is modified by adding to the bare action a regulator term (in the context of a scalar field $\varphi$) 
\be
\Delta S_{k}\left[ \varphi \right] =\frac{1}{2} \int \frac{d^{4}p}{\left( 2\pi \right)^{4}} \varphi \left( -p \right) R_{k}\left( p^{2} \right) \varphi \left( p \right),\ee where $R_{k}\left( p^{2} \right)$ acts like a momentum-dependent mass: for $p^{2}\ll k^2$, $R_{k}\left( p^{2} \right) \sim k^2$,suppresses IR modes; for $p^{2}\gg k^2$, $R_{k}\left( p^{2} \right) \to 0$ leaves UV modes untouched. The starting point of the FRG is a \emph{scale-dependent} generating function \cite{Wetterich:1992yh} with source $J$
\be Z_{k}\left[ J \right] =\int D\varphi \exp \left( -S\left[ \varphi \right] -\Delta S_{k}\left[ \varphi \right] +\int J\varphi \right),\ee and its Legendre transform, the effective average action,
\be\Gamma_{k} \left[ \Phi \right] =\sup_{J} \left\{ \int J\Phi -\log Z_{k}\left[ J \right] \right\} -\Delta S_{k}\left[ \Phi \right]\ee
with $\Phi =\left< \varphi \right>_{J}$.The evolution of $\Gamma_{k}$ with the FRG  ``time'' $t=\log k$ is governed by the Wetterich equation \eqref{Wetterich_1}.
Here $\Gamma_{k}^{\left( 2 \right)}$ denotes the second functional derivative (Hessian) with respect to all fields $\Phi$; $R_{k}$ is viewed as an operator  in field space (since $\Delta S_{k}\left[ \varphi \right]$ depends on $\varphi$). 
Although this formalism is in principle exact, there are difficulties in obtaining rigorous results using it. These difficulties are
\begin{itemize}
    \item scheme dependence such as the choice of the regulator function and gauge \cite{Toms:2010vy}
    \item the truncation of the terms used in the average effective action. The theories considered are perturbatively nonrenormalisable in 4-dimensions. In perturbative renormalisation an infinite number of terms with higher mass dimensions are required, making different truncations relevant upto different energy scales. The truncation of the terms in the average effective action is a reflection of this issue (present in the perturbative framework). 
\end{itemize}
 
\section{A toy model  for a scalar FRG and LPA$^\prime$}\label{sec:FRGscalar}
We consider a simple model for a single scalar field (in a flat Euclidean background). The scale-dependent effective average action is truncated to
\be
\Gamma_{k}[\phi]=\int d^{4} x\left[\frac{1}{2} Z_{k }\left(\partial_{\mu} \phi\right)^{2}+U_{k}(\phi)\right]
\label{eq:LPAprime_truncation}
\ee
where \(U_k(\phi)\) is the scale-dependent effective potential and \(Z_k\) is a
scale-dependent, field-independent wave-function renormalisation
and fluctuations $\delta \phi(x)$($=\int \frac{d^{4} p}{(2 n)^{4}} \delta \tilde{\phi}(p) e^{i p \cdot x}$)   around a constant background $\bar{\phi}$. This truncation corresponds to the LPA$^\prime$ approximation, i.e. the leading
order of the derivative expansion augmented by a running kinetic prefactor.
 Using both sides of  the Wetterich equation (\ref{Wetterich_1}), we calculate $\frac{\partial}{\partial{t}} Z_{k \phi}$ and $ \partial_{t} U_{k}$ in terms of $t={\rm{log}} \tfrac{k}{\Lambda}$ (where $\Lambda$ is an appropriate UV cut-off), which we then equate; this is the general procedure used in other models and allows calculations of beta functions and anomalous dimensions \cite{Peskin:1995ev}. 
 In LPA$^\prime$, the infrared regulator is chosen to incorporate the running
wave--function renormalisation \(Z_k\),
\begin{equation}
R_k(p^2)
=
Z_k\,k^2\,r\!\left(\frac{p^2}{k^2}\right),
\label{eq:LPAprime_regulator}
\end{equation}
where \(r(y)\) is a dimensionless shape function. A convenient choice is the
Litim regulator,
\begin{equation}
r(y)=\left(\frac{1}{y}-1\right)\Theta(1-y).
\end{equation}
With this choice, the scale derivative of the regulator reads
\begin{equation}
\partial_t R_k(p^2)
=
Z_k\,k^2
\left[
(2-\eta)\,r(y)-2y\,r'(y)
\right],
\qquad
\eta \equiv -\partial_t\ln Z_k .
\label{eq:dRk}
\end{equation}
For a constant background field \(\bar\phi\), the second functional derivative of
the effective action is
\be
\frac{\delta^{2} \Gamma_{k}}{\delta \phi(x) \delta \phi(y)}=\Gamma_{k}^{(2)}(x,y)=\left(-Z_{k} \partial_{x}^{2}+U_{\cy k \cbl}^{\prime \prime}(\bar{\phi})\right) \delta^{(4)}(x-y),
\ee
and we find
\be
\label{LPA}
\partial_{t} U_{k}(\bar{\phi})=\frac{1}{2} \int \frac{d^{4} q}{(2 \pi)^{4}} \frac{\partial_{t} R_{k}\left(q^{2}\right)}{Z_{k} q^{2}+R_{k}\left(q^{2}\right)+U_{k}^{\prime \prime}(\bar{\phi})}=\cy 
\frac{k^4}{2} l_{0}^{4}(w) \,,
\ee
where
\be
\cy l_{0}^{4}(w)=\frac{1}{k^{4}} \int \frac{d^{4} q}{(2 \pi)^{4}} \frac{\partial_{t} R_{k}\left(q^{2}\right) / Z_{k}}{q^{2}+\frac{R_{k}\left(q^{2}\right)}{Z_{k}}+k^{2} w} \,,
\ee
and $w=\frac{U_{k}^{\prime \prime}(\phi)}{Z_{k} k^{2}}$. $l_{0}^{4}$ is an example of a threshold function, which is dependent on the regulator function. In terms of a physical mass $m^{2}=k^{2} w$, the denominator of the integrand of the threshold function is dominated by $q^{2}+\frac{R_{k}\left(q^{2}\right)}{Z_{k}}$ for $\frac{m^{2}}{k^{2}} \ll 1$. For $\frac{m^{2}}{k^{2}} \gg 1$, $l_{0}^{4}(w)$ is small, since the $m^2$ term in the denominator is dominant. This shows why a function such as $l_{0}^{4}$ is called a threshold function.

From the Wetterich equation\cy, if we include the change of wavefunction renormalization with the energy scale
\cbl
\be
\partial_{t} Z_{k}=\lim _{p^{2} \rightarrow 0} \frac{\partial}{\partial {{p}^{2}}} \partial_{t} \Gamma^{(2)}(p).
\ee
which gives the anomalous dimension through $\eta =-\dfrac{\partial }{\partial \cy t\cbl}\log Z_{k}$.
Using the right hand side of the Wetterich equation we obtain, in momentum space:
\be
\partial _{t}\Gamma _{k}^{\left( 2\right)}(p,-p) =\int {d^4}q  U'''_k(\overline{\phi })^2 G(q^2)^2\frac{\partial^2 G((p-q)^2)}{\partial p^2}     \partial_t R_k (q^2) \,,
\ee
where $U'''_k(\overline{\phi })$ is the third functional derivative with respect to the field $\phi$ evaluated at the constant $\overline{\phi }$ and  $ G(p^2) = \frac{ 1}{Z_k p^2 + U''_k(\overline{\phi }) +r_k(p^2)}$. 

After some algebra, we obtain 
\be
\partial _{t}Z_{k}=v_{4}k^{4}l_{1}^{4}\left( w\right) \left( U_{k}'''\left( \overline{\phi }\right) \right) ^{2}
\ee
where
\be
l_{1}^{4}\left( w\right) =\dfrac{1}{2v_{4}k^{4}}\int \dfrac{d^{4}q}{\left( 2\pi \right) ^{4}}\dfrac{(G'+2p^2G'')\partial _{t}R_{k}\left( q^{2}\right) }{\left( Z_{k}q^{2}+R_{k}\left( q^{2}\right) + Z_{k} k^{2}w\right) ^{2}}
\ee
\cbl
is another threshold function, where  $v_d$ denotes the d-dimensional angular volume factor. We have shown a typical calculation using FRG. In general it is necessary to choose a $\Gamma_k$ which may have gauge fields; a decision is made on the degrees of freedom relevant for the physical situation at hand and this may lead to an introduction of new fields (e.g. the metric \footnote{We shall primarily consider the background expansion 
\eqref{backexp} about 
a flat (Minkowski) background metric $\overline g_{\mu\nu}=\eta_{\mu\nu}$.} $g_{\mu\nu}$, with $g=\det g_{\mu \nu }$) and interactions (e.g. higher derivative terms) 
\begin{align}
\Gamma _{k}\left[ \,\phi ,g\right] &=\int d^{4}x\sqrt{g}\,\Big[ \,\dfrac{1}{2}Z_{\phi ,k}\phi \left( -\nabla ^{2}\right) \phi +\dfrac{1}{2}\overline{\rho}_{a,k}\left( g^{\mu \nu }\partial _{\mu }\phi \partial _{\nu }\phi \right) ^{2}+\dfrac{1}{2}\overline{\rho}_{b,k}\left( g^{\mu \nu }\nabla _{\mu }\partial _{\nu }\phi \right) ^{2} \nonumber \\
&+\dfrac{1}{2}\overline{\rho}_{c,k}\left( g^{\mu \nu }\partial _{r}\phi \partial _{\nu }\phi \right) \nabla ^{2}\phi +U\left( \phi \right) \Big]. 
\end{align}
For a more general field theory with $n$ fields $\Phi_j$ ($j=1, \cdots, n$), choose the standard LPA$^\prime$ regulator
\begin{equation}
R_{k,j}(p^2)=Z_j(k)\,k^2\,r_j(y).
\end{equation}

whre $y \equiv \frac{p^2}{k^2}$ Using \(\partial_t Z_j=-\eta_j Z_j\) and \(\partial_t y=-2y\), one finds
\begin{align}
\partial_t R_{k,j}(p^2)
&= (\partial_t Z_j)\,k^2 r_j(y)
   + Z_j\,\partial_t\!\big(k^2 r_j(y)\big)
\nonumber\\[4pt]
&= Z_j(k)\,k^2
\Big[
(2-\eta_j)\,r_j(y)
-2y\,r_j'(y)
\Big].
\label{Inhomogeneous}
\end{align}
Starting from the Wetterich equation
\begin{equation}
\partial_t \Gamma_k
=
\frac{1}{2}\,\mathrm{STr}\!\left[
\big(\Gamma_k^{(2)}+R_k\big)^{-1}\,
\partial_t R_k
\right],
\end{equation}
the exact flow of the two-point function for a field \(i\) is obtained by
two functional derivatives with respect to \(\phi_i\).
One finds generically
\begin{equation}
\label{MasterEq}
\partial_t \Gamma^{(2)}_{k,ii}
=
\frac{1}{2}\,
\mathrm{STr}\!\left[
G_k\,\Gamma^{(3)}_{k,i}\,
G_k\,\Gamma^{(3)}_{k,i}\,
G_k\,\partial_t R_k
\right]
-
\frac{1}{2}\,
\mathrm{STr}\!\left[
G_k\,\Gamma^{(4)}_{k,ii}\,
G_k\,\partial_t R_k
\right]
\end{equation}
where
\begin{equation}
\label{MasterEquation}
G_k \equiv \big(\Gamma_k^{(2)}+R_k\big)^{-1}
\end{equation}
is the full regularised propagator, and \(\mathrm{STr}\) denotes a supertrace
over momentum, internal indices, and statistics. Some of these indices are suppressed in \eqref{MasterEq} and  \eqref{MasterEquation}. 

Explicitly, the first term on the right of \eqref{MasterEq}, the rainbow (sunset) contribution, reads
\begin{align}
\partial_t \Gamma^{(2)}_{k,ii}\big|_{\mathrm{rainbow}}
&=
\frac{1}{2}
\sum_{a,b,c,d}
\int\!\frac{d^d q}{(2\pi)^d}\;
(G_k)_{ab}(q)\,
(\Gamma^{(3)}_{k,i})_{bc}(p;q,-q-p)\,
(G_k)_{cd}(q+p)
\nonumber\\
&\qquad\times
(\Gamma^{(3)}_{k,i})_{da}(-p;q+p,-q)\,
(\partial_t R_k)_{aa}(q),
\end{align}
and the second term, the tadpole contribution, is
\begin{align}
\partial_t \Gamma^{(2)}_{k,ii}\big|_{\mathrm{tadpole}}
&=
-\frac{1}{2}
\sum_{a,b}
\int\!\frac{d^d q}{(2\pi)^d}\;
(G_k)_{ab}(q)\,
(\Gamma^{(4)}_{k,ii})_{ba}(p,-p;q,-q)\,
(\partial_t R_k)_{aa}(q).
\end{align}
The regulator insertion \(\partial_t R_k\) always carries the field index
of the internal loop propagator, which is why anomalous dimensions enter
the LPA$^\prime$ equations exclusively through the combinations
\((2-\eta_j)\) for the fields \(j\) circulating in the loop.

 For theories with local invariance, it is necessary to choose a gauge, and introduce any necessary Faddev-Popov ghosts~\cite{Peskin:1995ev}. Given  the difficulties of implementing rigorously  quantum  diffeomorphism  invariance~\cite{Falls:2017lst} in the study of gravity, our work on the axion model will be  suggestive (but not conclusive). 

\section{ Differential forms and the average action}
\label{DiffForms}

At a point $x \in M$, the spacetime manifold, the cotangent space $T^*_x M$ is the vector space of all linear maps of the tangent space $T_x M$ to $\mathbb R$. A $n$ form $\omega$ at $x$ is an alternating multilinear form \cite{Nakahara2018-on}

\be \omega:T_{x}M\times \ldots \times T_{x}M\rightarrow \mathbb R.\ee

The space of all such $n$ forms is denoted by $\Omega^n(M)$. A gauge field $A \in \Omega^1(M)$ and might be Lie algebra valued. The field strength $F \in \Omega^2(M)$ is given by 
\be F=dA+A\wedge A ,\ee 
with $A\wedge A$ absent for an Abelian gauge theory. The scalar axion field $b \in \Omega^0(M)$; the metric fluctuation $h\in \Gamma \left( T^{\ast }M\otimes T^{\ast }M\right)  $ (the space of smooth sections on $T^{\ast }M\otimes T^{\ast }M$) where the metric $g_{\mu \nu }=\eta _{\mu \nu }+\kappa h_{\mu \nu }$, and $\eta _{\mu \nu }$ is the flat background.

We use the Hodge star, which takes a $p$-form on an oriented $d$-dimensional Riemannian manifold and produces a $(d-p)$ form. As an example, for the 2-form field tensor
\be F=\dfrac{1}{2}F_{\mu \nu }dx^{\mu }\wedge {dx^{\nu }}\,, 
\ee
\be
\ast F=\dfrac{1}{2}\left( \ast F\right)_{\rho\sigma }dx^{\rho }\wedge {dx^{\sigma }}\,,\ee
and
\be
\left( \ast F\right) _{\rho\sigma }=\dfrac{1}{2}\sqrt{\lvert g \rvert }\varepsilon_{\rho \sigma}^{\quad\mu \nu }F_{\mu \nu}\,.
\ee
Also 
\be \int F{\wedge }F=\dfrac{1}{4}\int d^{4}x\varepsilon ^{\mu \nu \rho \sigma }F_{\mu \nu }F_{\rho\sigma} \ee and 
\be
\int F \wedge\ast F= \dfrac{1}{2} F_{\mu \nu }F^{\mu \nu }\sqrt{\left| g\right| }d^{4}x.\ee
The form notation elegantly takes into account metric factors  and is convenient when discussing topology.

\section{Super--diagram rules for the FRG}\label{sec:appB}

We will outline compact `Feynman-rule' set for the Wetterich equation, which can be written as 
\begin{align}
  \partial_t\Gamma_k
  \;=\;
  \tfrac12\,
  \mathrm{STr}\,
  \bigl[
      (\Gamma^{(2)}_k + R_k)^{-1}
      \,\partial_t R_k
  \bigr]
  \;=\;
  \tfrac12\,
  \mathrm{STr}\,
  \bigl[
      (\underbrace{\mathcal P_k}_{\text{propagator}}
      +\underbrace{\mathcal I_k}_{\text{interaction}})^{-1}
      \,\partial_t R_k
  \bigr].
\end{align}
This can be unpacked into super-diagrams built from 
\begin{itemize}
    \item a dressed propagator $G_{k}\equiv P_{k}^{-1}$ (block-diagonal in York basis for gravity)
    \item an interaction insertion $I_k$ (which carries no derivatives, but all other types of vertices)
\end{itemize}
\begin{align}
  \mathcal P_k :=
     \Gamma^{(2)}_k\big|_{\text{derivatives}}
     + R_k\!\bigl(-\bar\nabla^{2}\bigr),
  \qquad
  \mathcal I_k :=
     \Gamma^{(2)}_k -
     \Gamma^{(2)}_k\big|_{\text{derivatives}}.
\end{align}
We will now give a geometric series expansion.
Denote $G_k := \mathcal P_k^{-1}$ and so  then
\begin{align}
  (\mathcal P_k + \mathcal I_k)^{-1}
  &= G_k\,
     \sum_{n=0}^{\infty}
     \bigl(-\mathcal I_k G_k\bigr)^{n}, 
 \\[4pt]
 \partial_t\Gamma_k
  &= \frac12
     \sum_{n=0}^{\infty}
     (-1)^{\,n}\;
     \mathrm{STr}\,\Bigl[
       \partial_t R_k\,
       G_k\,
       (\mathcal I_k G_k)^{\,n}
     \Bigr].
\end{align}
With the above expansion every term can be drawn as a (one--loop) ``super--diagram'' according to the rules: $ G_k \;=\;\mathcal P_k^{-1}$
        (using the appropriate York block)
which can be represented by an internal line, a dot can represent a vertex $-\mathcal I_k$ which  depends on the background curvature; cross  represents $\partial_t R_k\bigl(-\bar\nabla^{2}\bigr)$, (exactly \emph{one} per loop); there is an integration over momentum and sum over indices; STr is $+$ for bosons and $-$ for ghosts and there is prefactor of $1/2$ for every closed loop.

\vspace{.6em}
\noindent
Here are examples:
\begin{align}
  n=0\!:~~
  \partial_t\Gamma_k^{(0)} &=
  \tfrac12\,\mathrm{STr}\,\bigl[\partial_t R_k\,G_k\bigr]
  &&\text{(tadpole)}
\\[2pt]
  n=1\!:~~
  \partial_t\Gamma_k^{(1)} &=
  -\tfrac12\,\mathrm{STr}\,\bigl[\partial_t R_k\,G_k\,\mathcal I_k\,G_k\bigr]
  &&\text{(one--vertex loop)}
\end{align}

\section{Flat–space York (TT) decomposition and Barnes–Rivers projectors}
\label{FlatYork}

\subsection{Flat–space York/TT decomposition}
On a flat $d$-dimensional spacetime background $\eta_{\mu\nu}$ (Euclidean or Minkowski), any symmetric tensor
$h_{\mu\nu}(x)$ admits the decomposition
\begin{equation}
h_{\mu\nu}
= h^{\mathrm{TT}}_{\mu\nu}
+ \partial_{(\mu}\xi^{\;T}_{\nu)}
+ \Big(\partial_\mu\partial_\nu-\frac{1}{d}\,\eta_{\mu\nu}\,\partial^2\Big)\sigma
+ \frac{1}{d}\,\eta_{\mu\nu}\,\varphi\,,
\label{eq:flatYork}
\end{equation}
with constraints
\begin{equation}
\partial^\mu h^{\mathrm{TT}}_{\mu\nu}=0,\qquad
\eta^{\mu\nu}h^{\mathrm{TT}}_{\mu\nu}=0,\qquad
\partial^\mu\xi^{\;T}_\mu=0,\qquad
\varphi\equiv \eta^{\rho\sigma}h_{\rho\sigma}.
\end{equation}
A useful gauge–invariant scalar is
\begin{equation}
\chi \;\equiv\; \varphi - \partial^2\sigma\quad\Rightarrow\quad \delta\chi=0
\ \ \text{under}\ \ \delta h_{\mu\nu}=\partial_\mu\epsilon_\nu+\partial_\nu\epsilon_\mu.
\end{equation}

\subsection{ Momentum–space inversion (practical formulas)}
In Fourier space ($\partial_\mu\!\to\! i p_\mu$), define
\begin{equation}
\theta_{\mu\nu}(p)=\eta_{\mu\nu}-\frac{p_\mu p_\nu}{p^2},\qquad
\omega_{\mu\nu}(p)=\frac{p_\mu p_\nu}{p^2}.
\end{equation}
Given $h_{\mu\nu}(p)$, the York pieces are
\begin{align}
\varphi(p) &= \eta^{\mu\nu}h_{\mu\nu}(p),\\
\sigma(p) &= \frac{d}{d-1}\,\frac{1}{p^2}\!\left[
\frac{p_\mu p_\nu}{p^2}\,h^{\mu\nu}(p)-\frac{1}{d}\,\varphi(p)\right],\\
\xi^{\;T}_\nu(p) &= \frac{2}{p^2}\,\theta_{\nu}{}^{\alpha}(p)\,p^\mu h_{\mu\alpha}(p),\\
h^{\mathrm{TT}}_{\mu\nu}(p) &=
h_{\mu\nu}(p)-\frac12\!\left(p_\mu \xi^{\;T}_\nu+p_\nu \xi^{\;T}_\mu\right)
-\Big(p_\mu p_\nu-\tfrac{1}{d}\eta_{\mu\nu}p^2\Big)\sigma(p)
-\tfrac{1}{d}\,\eta_{\mu\nu}\,\varphi(p).
\end{align}

\subsection{ Barnes–Rivers spin projectors (momentum space)}
Projectors on the symmetric rank-2 subspace (for $p^2\neq0$):
\begin{align}
P^{(2)}_{\mu\nu\rho\sigma} &=
\frac12\!\left(\theta_{\mu\rho}\theta_{\nu\sigma}+\theta_{\mu\sigma}\theta_{\nu\rho}\right)
-\frac{1}{d-1}\,\theta_{\mu\nu}\theta_{\rho\sigma},\\[2pt]
P^{(1)}_{\mu\nu\rho\sigma} &=
\frac12\!\left(\theta_{\mu\rho}\omega_{\nu\sigma}
+\theta_{\mu\sigma}\omega_{\nu\rho}
+\theta_{\nu\rho}\omega_{\mu\sigma}
+\theta_{\nu\sigma}\omega_{\mu\rho}\right),\\[2pt]
P^{(0\!-\!s)}_{\mu\nu\rho\sigma} &= \frac{1}{d-1}\,\theta_{\mu\nu}\theta_{\rho\sigma},\\[2pt]
P^{(0\!-\!w)}_{\mu\nu\rho\sigma} &= \omega_{\mu\nu}\,\omega_{\rho\sigma}.
\end{align}
Useful “bridge” operators in the scalar sector:
\begin{equation}
P^{(0\!-\!sw)}_{\mu\nu\rho\sigma}=\frac{1}{\sqrt{d-1}}\,\theta_{\mu\nu}\omega_{\rho\sigma},\qquad
P^{(0\!-\!ws)}_{\mu\nu\rho\sigma}=\frac{1}{\sqrt{d-1}}\,\omega_{\mu\nu}\theta_{\rho\sigma}.
\end{equation}

\noindent\textbf{Algebra and completeness:}
\begin{align}
&P^{(A)}P^{(B)}=\delta^{AB}P^{(B)}\quad (A,B\in\{2,1,0\!-\!s,0\!-\!w\}),\\
&P^{(0\!-\!sw)}P^{(0\!-\!ws)}=P^{(0\!-\!s)},\quad
P^{(0\!-\!ws)}P^{(0\!-\!sw)}=P^{(0\!-\!w)},\\
&\mathbb{I}_{\mu\nu\rho\sigma}=\tfrac12(\eta_{\mu\rho}\eta_{\nu\sigma}+\eta_{\mu\sigma}\eta_{\nu\rho})
= P^{(2)}+P^{(1)}+P^{(0\!-\!s)}+P^{(0\!-\!w)}.
\end{align}

\subsection{ One-line use case (spin-2 projection)}
For a two-point kernel $\Gamma^{(2)}_{hh}(p)$, the spin-2 kinetic coefficient can be read off without a York split via
\begin{equation}
Z_N(k)=\frac{1}{\mathrm{tr}\,P^{(2)}}\,
\frac{\partial}{\partial p^2}\Big[\,P^{(2)}(p)\!:\!\Gamma^{(2)}_{hh}(p)\,\Big]\Big|_{p^2=k^2},
\qquad A\!:\!B\equiv A_{\mu\nu\rho\sigma}B^{\mu\nu\rho\sigma}.
\end{equation}
(For $d=4$, $\mathrm{tr}\,P^{(2)}=5$.)
\cbl
\section{Details of loop diagrams required for the calculations of anomalous dimensions} 
\label{LoopDetails}

 To obtain the anomalous dimensions themselves, one differentiates the flow equation 
twice with respect to the fields and projects onto the kinetic structures. Because 
the only interactions present are the axion--photon CS vertex, minimal gravitational 
couplings, and (optionally) the axion--gravity CS vertex, the anomalous dimensions 
receive a finite set of \emph{one-loop} contributions. The truncation we adopt keeps all operators that are generated by one-graviton or 
one-gauge-boson exchange at the order relevant for the renormalisation of the Chern-Simons couplings.
\begin{align}
\big[\partial_t \Gamma^{(2)}_{bb}(p)\big]_{A\text{-loop}}
&= +\frac{1}{2}\!\int\!\frac{d^dq}{(2\pi)^d}\;
V_{bAA}^{\mu\nu}(p,q,-p{-}q)\;
G^{A}_{\mu\mu'}(q)\;
\partial_t R_A(q)\;
G^{A}_{\nu\nu'}(p{+}q)\;
V_{bAA}^{\mu'\nu'}(-p,-q,p{+}q).
\label{eq:AxionPhotonLoop}
\end{align}
Here:
\be
G^{A}_{\mu\nu}(q)=\frac{T_{\mu\nu}(q)}{Z_A\,[\,q^2+R_A(q)\,]},
\qquad
T_{\mu\nu}(q)=\eta_{\mu\nu}-\frac{q_\mu q_\nu}{q^2},
\ee
\be
\partial_t R_X(q)=Z_X\Big[(2-\eta_X)\,k^2\,r(y)-2y\,k^2\,r'(y)\Big],\qquad y\equiv q^2/k^2,\quad X\in\{a,A,h\}.
\ee
and for the CS vertex (schematically, up to total antisymmetry)
\be
V_{bAA}^{\mu\nu}(p,q,-p{-}q)=g_{\mathrm{mix}}\;\epsilon^{\mu\nu\rho\sigma}\,q_\rho\,(p{+}q)_\sigma.
\ee
The axion anomalous dimension follows from projecting the \(p^2\)-term:
\be
\eta_b\;=\;-\partial_t\ln Z_b
\;\propto\;
\left.\frac{\partial}{\partial p^2}\big[\partial_t \Gamma^{(2)}_{bb}(p)\big]_{A\text{-loop}}\right|_{p^2=k^2}.
\ee
This contribution scales as \(\eta_b\sim \mathcal O(g_{\mathrm{mix}}^2)\) (modulo threshold factors).
\vskip .1cm
\noindent Similarly, for completeness we give the graviton loop contribution from an axion--gravity CS vertex (although we shall omit it for phenomenological reasons)  
\begin{align}
\big[\partial_t \Gamma^{(2)}_{bb}(p)\big]_{h\text{-loop}}
&= +\frac{1}{2}\!\int\!\frac{d^dq}{(2\pi)^d}\;
V_{b hh}^{\alpha\beta\rho\sigma}(p,q,-p{-}q)\;
G^{h}_{\alpha\beta\mu\nu}(q)\;
\partial_t R_h(q)\;
G^{h}_{\rho\sigma\kappa\lambda}(p{+}q)\;
V_{b hh}^{\mu\nu\kappa\lambda}(-p,-q,p{+}q).
\label{eq:AxionGravLoop}
\end{align}
With TT projection:
\be
G^{h}_{\alpha\beta\rho\sigma}(q)=\frac{P^{(2)}_{\alpha\beta\rho\sigma}(q)}{Z_h\,[\,q^2+R_h(q)\,]},
\qquad
\partial_t R_h(q)=Z_h\Big[(2-\eta_h)\,k^2\,r\!\left(\tfrac{q^2}{k^2}\right)+\cdots\Big],
\ee
\be
P^{(2)}_{\alpha\beta\rho\sigma}=\tfrac12\!\left(\theta_{\alpha\rho}\theta_{\beta\sigma}+\theta_{\alpha\sigma}\theta_{\beta\rho}\right)
-\tfrac{1}{3}\theta_{\alpha\beta}\theta_{\rho\sigma},\quad
\theta_{\mu\nu}=\eta_{\mu\nu}-\frac{q_\mu q_\nu}{q^2}.
\ee
\begin{align}
V_{b hh}^{\mu\nu\rho\sigma}(p_a,k,\ell)
&= i g_{\rm CS}\,\epsilon^{\alpha\beta\gamma\delta}\,
\mathcal{R}_{\alpha\beta}{}^{\mu\nu}(k)\,
\mathcal{R}_{\gamma\delta}{}^{\rho\sigma}(\ell)
\;+\;(\mu\nu,k)\leftrightarrow(\rho\sigma,\ell),\\
\cy
\mathcal{R}_{\alpha\beta\mu\nu}(k)
&=\tfrac12\Big(k_\alpha k_\mu h_{\beta\nu}-k_\alpha k_\nu h_{\beta\mu}
                -k_\beta  k_\mu h_{\alpha\nu}+k_\beta  k_\nu h_{\alpha\mu}\Big).
\end{align}
The CS vertex \(V_{a hh}\) comes from expanding \(a\,R\tilde R\) to quadratic order in \(h\); it carries a coupling \(g_{\mathrm{CS}}\) and momenta contracted with the Levi–Civita tensor. Then
\be
\eta_b\;\propto\;
\left.\frac{\partial}{\partial p^2}\big[\partial_t \Gamma^{(2)}_{bb}(p)\big]_{h\text{-loop}}\right|_{p^2=k^2}
\;\sim\; g_{\mathrm{CS}}^2 \times \big(\text{thresholds involving }g_N,\,\eta_h\big).
\ee
Similarly for the photon anomalous dimension $\eta_A$ we have
\begin{align}
\big[\partial_t \Gamma^{(2)}_{AA}(p)\big]_{b\text{-loop}}
&=\frac{1}{2}\!\int\!\frac{d^4 q}{(2\pi)^4}\;
V^{\mu\alpha}_{{AA}b}(p,q,-p{-}q)\,
G_b(q)\,\partial_t R_b(q)\,G_b(p{+}q)\,
V^{\nu}{}_{\alpha\,{bA}}(-p,-q,p{+}q),\\
\big[\partial_t \Gamma^{(2)}_{AA}(p)\big]_{h\text{-loop}}
&=\frac{1}{2}\!\int\!\frac{d^4 q}{(2\pi)^4}\;
V^{\mu\alpha|\rho\sigma}_{AAh}(p,q,-p{-}q)\,
G^h_{\rho\sigma,\rho'\sigma'}(q)\,\partial_t R_h(q)\,
G^h_{\alpha\beta,\alpha'\beta'}(p{+}q)\,
V^{\nu\beta|\rho'\sigma'}_{AAh}(-p,-q,p{+}q).
\end{align}
where we denote the propagators by
\be
G_a(q)=\frac{1}{Z_b\,[q^2+R_a(q)]},\quad
G^A_{\alpha\beta}(q)=\frac{T_{\alpha\beta}(q)}{Z_A\,[q^2+R_A(q)]},\quad
G^h_{\mu\nu,\rho\sigma}(q)=\frac{P^{(2)}_{\mu\nu,\rho\sigma}(q)}{Z_h\,[q^2+R_h(q)]}.
\ee
In our one-metric approximation $Z_{h}$ is identified with $Z_N$.
The TT graviton anomalous dimension is calculated from 
\be
\eta_h=-\frac{1}{Z_h}\left.\frac{\partial}{\partial p^2}
\left[
\frac{1}{5}\,P^{(2)}_{\mu\nu\rho\sigma}(p)\,\partial_t\Gamma^{(2)\,\mu\nu\rho\sigma}_{hh}(p)
\right]\right|_{p^2=k^2}.
\ee
with matter contributions:
\begin{align}
\label{etagraviton}
\big[\partial_t \Gamma^{(2)}_{hh}(p)\big]_{A\text{-loop}}
&=\frac{1}{2}\!\int\!\frac{d^4 q}{(2\pi)^4}\;
V^{\mu\nu|\alpha\beta}_{hAA}(p,q,-p{-}q)\,
G^A_{\alpha\alpha'}(q)\,\partial_t R_A(q)\,
G^A_{\beta\beta'}(p{+}q)\,
V^{\rho\sigma|\alpha'\beta'}_{hAA}(-p,-q,p{+}q),\\
\big[\partial_t \Gamma^{(2)}_{hh}(p)\big]_{b\text{-loop}}
&=\frac{1}{2}\!\int\!\frac{d^4 q}{(2\pi)^4}\;
V^{\mu\nu}_{hbb}(p,q,-p{-}q)\,
G_b(q)\,\partial_t R_b(q)\,
G_b(p{+}q)\,
V^{\rho\sigma}_{hbb}(-p,-q,p{+}q),
\end{align}
where the relevant vertices are \be
V^{\mu\nu|\rho\sigma}_{hAA}(p,q,-p{-}q)
=\frac{\kappa}{2}\;\{
\begin{aligned}[t]
&\delta_{\rho\sigma}\,(p_\mu q_\nu + p_\nu q_\mu - \delta_{\mu\nu}\,p\!\cdot\! q) \\
&-\,\delta_{\mu\rho}\,(q_\nu p_\sigma + q_\sigma p_\nu - \delta_{\nu\sigma}\,p\!\cdot\! q)
    -\,\delta_{\nu\rho}\,(q_\mu p_\sigma + q_\sigma p_\mu - \delta_{\mu\sigma}\,p\!\cdot\! q) \\
&-\,\delta_{\mu\sigma}\,(p_\nu q_\rho + p_\rho q_\nu - \delta_{\nu\rho}\,p\!\cdot\! q)
    -\,\delta_{\nu\sigma}\,(p_\mu q_\rho + p_\rho q_\mu - \delta_{\mu\rho}\,p\!\cdot\! q) \\
&+\,\delta_{\mu\nu}\,(p_\rho q_\sigma + p_\sigma q_\rho - \delta_{\rho\sigma}\,p\!\cdot\! q)\}\,.
\end{aligned}\ee
and 
\be
V^{\mu\nu}_{hbb}(-p,-q,p{+}q)
= \frac{\kappa\,Z_b}{2}\,
\Big[\, p_\mu q_\nu + p_\nu q_\mu \;-\; \delta_{\mu\nu}\,\big(p\!\cdot\! q + m_b^2\big) \Big].\ee

These are the ingredients that go into the LPA$'$. The evaluation of the \emph{loop integrals} can be expressed in terms of \emph{threshold functions}
\be
\Phi_n^p(w)\equiv \frac{1}{\Gamma(n)}\!\int_0^\infty\!dy\,y^{\,n-1}
\frac{r(y)-y r'(y)}{\big[y(1+r(y))+w\big]^p},
\qquad
\tilde\Phi_n^p(w)\equiv \frac{1}{\Gamma(n)}\!\int_0^\infty\!dy\,y^{\,n-1}
\frac{r(y)}{\big[y(1+r(y))+w\big]^p}.
\ee
This type of derivation is typical and used in \cite{Dona:2015tnf,Reuter:2019byg}. However there are explicit tensorial structures (cf \eqref{etagraviton}) which appear here and have not been evaluated explcitly in published work. It is common to factor the regulator dependence entirely into $\Phi$ and $\tilde\Phi$ and present the remaining tensor coefficients as simple constants of $O(1)$: $\mathcal{A}^{(i)}$ (axion-photon projection coefficients), $\mathcal{G}^{(1)}$,$\tilde{\mathcal{G}}^{(1)}$ (from spin–2/0 graviton and ghost loops),\(\mathcal{M}_{S,V}(\lambda),\tilde{\mathcal{M}}_{S,V}(\lambda)\) quantify the back–reaction of minimally coupled scalar/vector matter on \(\eta_h\) .

\section{LPA$^\prime$ Anomalous Dimensions and Validity of the Simplified Approximation}
\label{LPAModified}

We consider an LPA$^\prime$ truncation of axion electrodynamics coupled to gravity,
keeping running wave--function renormalisations but neglecting higher--derivative
operators. The truncation is assumed to preserve axionic shift symmetry, {\it i.e.}
the axion has no nontrivial potential,
\begin{equation}
V(b)=0,
\end{equation}
so that the local axion--photon Chern--Simons coupling is not renormalised directly,
and its scale dependence enters only through wave--function renormalisations.

We define the dimensionless couplings
\begin{equation}
g_{b\gamma}^2 \equiv
\frac{g_{\rm mix,k}^2\,k^2}{Z_A^2 Z_b},
\qquad
g_N \equiv k^2 G_k,
\qquad
\lambda \equiv \frac{\Lambda_k}{k^2},
\end{equation}
and the anomalous dimensions ({\it cf.} \eqref{nbnfnh}):
\begin{equation}
\eta_b=-\partial_t\ln Z_b,
\qquad
\eta_A=-\partial_t\ln Z_A,
\qquad
\eta_N=-\partial_t\ln Z_h.
\end{equation}

\subsubsection*{General LPA$^\prime$ structure}

At the level of LPA$^\prime$, the anomalous dimensions receive contributions from
axion--photon loops and from gravitational dressing. Keeping only the correct
coupling dependence and leaving regulator--dependent numerical coefficients
unspecified, the general structure in a linearised framework is
\begin{align}
\eta_b &=
c_{bA}\, g_{b\gamma}^2
\;+\;
c_{bN}\, g_N
\;+\;
c_{bNA}\, g_N g_{b\gamma}^2,
\label{eq:eta_a_general}
\\[4pt]
\eta_A &=
c_{Ab}\, g_{b\gamma}^2
\;+\;
c_{AN}\, g_N
\;+\;
c_{ANA}\, g_N g_{b\gamma}^2,
\label{eq:eta_A_general}
\\[4pt]
\eta_N &=
c_{N}\, g_N
\;+\;
c_{NA}\, g_N g_{b\gamma}^2.
\label{eq:eta_N_general}
\end{align}
The coefficients \(c_i\) depend on the choice of regulator and gauge but are
expected to be \(\mathcal O(1)\). The terms proportional to \(g_{b\gamma}^2\)
encode axion--photon fluctuations; those proportional to \(g_N\) encode purely
gravitational dressing; and the mixed terms \(g_N g_{b\gamma}^2\) arise from
``interference'' diagrams coupling both sectors.

\subsubsection*{Self-consistency and the origin of singularities.}

In an LPA$'$ computation the anomalous dimensions are, however, not simply
polynomials in the couplings: they enter the scale derivative of the regulator
via factors such as \((2-\eta)\), so that \(\eta_b\) and \(\eta_A\) typically
appear on both sides of their defining flow equations. In the axion--photon
subsystem this leads to a \emph{linear self-consistency problem} of the form
\begin{equation}
\binom{\eta_b}{\eta_A}
=
\binom{\alpha_b}{\alpha_A}
+
U(g_{b\gamma}^2,g_N)\,\binom{\eta_b}{\eta_A},
\qquad
\text{with}\quad
\alpha_{b,A}=\mathcal O(g_{b\gamma}^2,g_N,g_N g_{b\gamma}^2),
\label{eq:selfconsistency_vector}
\end{equation}
This is a generalisation of the scalar case \eqref{eq:LPAprime_regulator} and \eqref{eq:dRk}. For the full system $U$ is a $3 \times 3$ matrix and the source column vector is $3$-dimensional.
In the LPA$'$ approximation the scale derivative of the regulator carries an
explicit dependence on the anomalous dimensions. For each field
$i\in\{a,A,N\}$ one introduces a regulator of the form
\begin{equation}
R_{k,i}(p^2)=Z_i(k)\,k^2\,r_i\!\left(\frac{p^2}{k^2}\right),
\end{equation}
where $Z_i(k)$ is the wavefunction renormalisation and $r_i(y)$ is a
dimensionless shape function. Taking the logarithmic scale derivative yields
\begin{equation}
\partial_t R_{k,i}(p^2)
=
Z_i k^2\!\left[
(2-\eta_i)\,r_i(y)
-
2y\,r_i'(y)
\right],
\qquad
\eta_i\equiv -\partial_t \ln Z_i .
\label{eq:dRk_eta}
\end{equation}
Consequently, every loop contribution to the Wetterich equation that involves
$\partial_t R_{k,i}$ splits naturally into a part proportional to the constant
term $2$ and a part proportional to $-\eta_i$. Upon projection onto the kinetic
operators that define the anomalous dimensions, the former generates
inhomogeneous contributions depending only on the couplings, while the latter
produces terms linear in the anomalous dimensions themselves. As a result, the
LPA$'$ flow equations for the anomalous dimensions take the generic
self--consistent linear form
\begin{equation}
\eta_i = s_i(\text{couplings}) + \sum_{j} U_{ij}(\text{couplings})\,\eta_j ,
\qquad i,j\in\{b,A,N\},
\end{equation}
where the coefficients $s_i$ and $U_{ij}$ are regulator- and gauge-dependent
threshold integrals.

\subsubsection*{Simplified approximation}
In a simplified treatment, we neglect mixed gravity--matter contributions
proportional to \(g_N g_{b\gamma}^2\), {\it i.e.} diagrams involving simultaneous
axion--photon and gravitational interactions. This corresponds to working
to linear order in \(g_N\) and linear order in \(g_{b\gamma}^2\), but not their product.
The system then reduces to
\begin{align}
\eta_b &=
\alpha\, g_{b\gamma}^2
\;+\;
\tilde\alpha\, g_N,
\label{eq:eta_a_simplified}
\\[4pt]
\eta_A &=
\beta\, g_{b\gamma}^2
\;+\;
\tilde\beta\, g_N,
\label{eq:eta_A_simplified}
\\[4pt]
\eta_N &=
\gamma\, g_N,
\label{eq:eta_N_simplified}
\end{align}
with \(\alpha,\tilde\alpha,\beta,\tilde\beta,\gamma=\mathcal{O}(1)\).

\subsubsection*{Justification for $\mathcal{O}(1)$ coefficients}

After rescaling loop momenta by the RG scale \(k\) and using an LPA$^\prime$
regulator of the form \(R_k=Z_k k^2 r(p^2/k^2)\), all FRG traces contributing
to anomalous dimensions reduce to dimensionless integrals multiplied by the
appropriate powers of the couplings. The remaining numerical factors arise
from angular integrals, projector algebra, and regulator threshold functions,
all of which are finite and typically of order unity away from threshold poles.
Therefore the coefficients in Eqs.~\eqref{eq:eta_a_general}--\eqref{eq:eta_N_general}
are naturally \(\mathcal{O}(1)\). The detailed amplitudes can be found in Appendix \ref{LoopDetails}.

\bibliographystyle{apsrev4-2}

\bibliography{CMSpaper.bib} 

\end{document}